\documentclass[a4paper,fleqn]{cas-dc}

\usepackage[numbers,sort&compress]{natbib}
\usepackage{amssymb}
\usepackage{bm}
\usepackage{cas-common}
\usepackage{tabularx}
\usepackage{amsmath}
\usepackage{hyperref}
\hypersetup{colorlinks=true,citecolor=blue,linkcolor=blue,urlcolor=blue}

\begin{document}
\let\WriteBookmarks\relax
\def\floatpagepagefraction{0.6}
\def\textpagefraction{0.1}
\shorttitle{Data-driven bath fitting for HD-DMFT}
\shortauthors{T. Kim et~al.}

\title[mode=title]{Data-driven bath fitting for Hamiltonian-diagonalization dynamical mean-field theory}

\author{Taeung Kim}
\author{Jeongmoo Lee}
\author[orcid=0000-0001-5497-5231]{Ara Go}
\cormark[1]
\ead{arago@jnu.ac.kr}

\cortext[1]{Corresponding author}
\label{sec:affiliations}

\affiliation{organization={Department of Physics, Chonnam National University},
      addressline={77 Yongbong-ro}, 
      city={Buk-gu},
      state={Gwangju},
      postcode={61186},
      country={South Korea}}

\begin{abstract}
We develop a machine-learning-based initialization strategy that alleviates the major practical bottleneck of Hamiltonian-diagonalization-based dynamical mean-field theory (HD-DMFT): nonlinear bath fitting. In HD-DMFT, the continuous hybridization function must be approximated by a finite set of bath-site energies and hybridization amplitudes, obtained by minimizing a highly non-convex multivariable cost function. As the number of bath sites increases, the optimization becomes increasingly sensitive to the initial guess and is prone to trapping in suboptimal local minima, which can substantially slow or destabilize the DMFT self-consistency loop. We recast bath fitting as a supervised regression problem and train a kernel ridge regression model to predict near-optimal discrete bath parameters directly from the target hybridization function on the Matsubara axis. A key methodological element is a physically grounded data-generation protocol: instead of random parameter sampling, we construct a diverse dataset from tight-binding Hamiltonians of layered-perovskite-like ruthenate models across systematically deformed structures, and obtain high-quality labels via fully converged conventional bath fitting. Time-reversal symmetry is incorporated explicitly in both feature and target representations to reduce effective dimensionality and enforce physical consistency. Benchmarks in the non-interacting limit show that the learned initialization systematically reduces the initial fitting error, decreases the number of conjugate-gradient iterations, and improves robustness against local minima over a wide range of bath sizes. Finally, we demonstrate transferability to an interacting DMFT calculation for Sr$_2$RuO$_4$ solved with an adaptive-truncation impurity solver, where the ML initialization yields consistently faster convergence than a symmetry-preserving heuristic baseline while preserving the final fitted solution.
\end{abstract}


\begin{NoHyper}
\maketitle
\end{NoHyper}
\label{sec:title}

\section{Introduction}
\label{sec:introduction}

Strong electronic correlations underlie a wide range of emergent phenomena in quantum materials, but they also render direct many-body simulations prohibitively expensive because the relevant Hilbert space grows exponentially with system size~\citep{Imada1998}.
Dynamical mean-field theory (DMFT) addresses this challenge by treating local dynamical correlations non-perturbatively while approximating nonlocal correlations at the mean-field level.
In DMFT, the infinite lattice problem is mapped onto a self-consistent quantum impurity model, in which an interacting impurity is embedded in an effective bath determined by the DMFT self-consistency condition.

In practical calculations, the performance of DMFT is largely determined by the choice of impurity solver.
Hamiltonian-diagonalization (HD)-based solvers are particularly attractive because they provide direct access to real-frequency quantities without analytic continuation and avoid statistical noise.
These approaches include exact diagonalization (ED)~\citep{Caffarel1994,Lu2017,Capone2007,Liebsch2011,Amaricci2022,Lorenzo2025} and truncated configuration-interaction (CI) variants~\citep{Zgid2011,Zgid2012,Lin2013,Go2015,Go2017,Herzog2025}.
However, HD-based solvers require discretizing the effective bath, and the number of orbitals that can be treated is ultimately limited by the exponential growth of the many-body Hilbert space.
Accordingly, the accuracy of the bath representation is constrained by the finite number of bath orbitals accessible in a given truncation strategy.

Substantial effort has therefore been devoted to increasing the accessible bath size, for instance through CI-based improvements~\citep{Go2017,Zgid2012} and other scalable representations such as matrix-product-state-based approaches~\citep{Nusspickel2020}. Because the impurity self-energy on the Matsubara axis typically converges relatively rapidly with increasing bath size, such advances can yield accurate approximations of the effective bath with a manageable number of bath orbitals~\citep{Senechal2010,MejutoZaera2020}.

Nevertheless, irrespective of the impurity solver employed, a central practical challenge remains: the bath-fitting procedure itself. Real-frequency impurity solvers that perform bath fitting directly on the real-frequency axis have also been proposed~\citep{Lu2014,Kitatani2023}; however, their systematic extension to realistic multi-orbital systems remains to be clarified.

In HD-DMFT, the continuous hybridization function must be represented by a finite set of discrete bath parameters (bath-site energies and hybridization strengths). Bath fitting is therefore formulated as the minimization of a nonlinear multivariable cost function measuring the mismatch between the target hybridization function and its discrete representation. The resulting optimization landscape is highly non-convex and typically contains many local minima. As the number of bath orbitals $N_{\mathrm{b}}$ increases, the dimension of the parameter space grows rapidly and the optimization becomes increasingly sensitive to the initial guess. If the initial bath is physically inappropriate (for example, it violates relevant symmetries or introduces energy scales outside the physically relevant window), the optimization can become trapped in a poor local minimum, which in turn may destabilize or even prevent convergence of the overall DMFT self-consistency loop~\citep{MejutoZaera2020,FlorezAblan2025}. In practice, this sensitivity often necessitates repeated fits with multiple initializations, increasing computational cost without guaranteeing convergence to the global optimum.

A variety of bath-fitting strategies have been proposed to mitigate these issues, including heuristic initialization rules and more systematic approaches that incorporate physical constraints or modify the optimization formulation~\citep{Koch2008,Senechal2010,MejutoZaera2020,Nusspickel2020,FlorezAblan2025,Foley2019}. Despite important progress, existing methods often remain computationally expensive, require problem-specific tuning, or provide inconsistent performance across diverse hybridization-function structures~\citep{Capone2004}. These limitations motivate the central question pursued here: can machine learning provide physically informed initial bath parameters that improve convergence robustness at minimal additional cost?

The need for robust bath discretization becomes especially acute in layered perovskite materials with complex low-energy electronic structures. Layered ruthenates are paradigmatic strongly correlated transition-metal oxides in which reduced dimensionality, crystal-field splitting, spin-orbit coupling, and substantial electron-electron interactions cooperate to generate exceptionally rich phase diagrams~\citep{Georges2013,Imada1998}. They exhibit metal-insulator transitions~\citep{Alexander1999}, diverse magnetic orderings~\citep{Braden1998}, heavy-fermion-like behavior~\citep{Nakatsuji2000}, and unconventional superconductivity~\citep{Maeno1994}. Prototypical Ruddlesden--Popper compounds such as Sr$_2$RuO$_4$ and Ca$_2$RuO$_4$ continue to attract sustained interest, including ongoing debates over pairing symmetry in Sr$_2$RuO$_4$\citep{Mackenzie2003,Maeno2024} and bandwidth-controlled Mott physics in Ca$_2$RuO$_4$\citep{Nakatsuji2000,Gorelov2010}. For these materials, obtaining a stable and accurate bath discretization is essential for reliable DMFT calculations.

In this work, we introduce a machine-learning-based initialization strategy to alleviate the difficulties associated with bath fitting.
Although the forward construction of the discrete hybridization function from a given set of bath parameters is straightforward, determining bath parameters that faithfully reproduce a lattice-derived target hybridization function is highly nontrivial.
In the DMFT context, the discrete bath representation is not a mere inversion of the hybridization function, but rather a constrained approximation of a continuous object that encodes lattice information.
Because a finite number of bath orbitals cannot reproduce the target hybridization function exactly, bath fitting becomes a nonlinear optimization problem in a high-dimensional parameter space.

In a previous study, we explored kernel ridge regression (KRR) as a
predictor of bath parameters from hybridization functions~\citep{Kim2024}. In that work,
training data were generated by a backward construction in which
discrete bath parameters $\{\varepsilon_l, V_l\}$ were first sampled
under prescribed protocols---ranging from regularly spaced bath energies
to fully random distributions---and the corresponding hybridization
functions were then evaluated through the discrete pole expansion
analogous to Eq.~\eqref{eq:discrete_hyb}. While the study confirmed
that KRR can learn the forward-consistent mapping, it also exposed a
key limitation: models trained on bath parameters with regular energy
distributions generalized successfully, whereas randomly distributed
bath energies failed to produce a generalizable model even with
training sets exceeding $10^{5}$ samples. The space of physically
relevant bath parameters is therefore too large and structured for
naive random sampling alone to suffice.

Here, we adopt a physically grounded data-generation strategy tailored to realistic multi-orbital systems.
We construct datasets from tight-binding Hamiltonians of layered ruthenate models across structurally deformed configurations and obtain high-quality training labels by performing conventional bath fitting to convergence for each case.
The resulting hybridization-function/bath-parameter pairs are used to train a KRR model that predicts near-optimal initial bath parameters for unseen hybridization functions. To ensure physical consistency and reduce the effective output dimensionality, we incorporate time-reversal symmetry directly into the feature design and the parameter representation.
By providing starting points closer to favorable basins of attraction in the cost-function landscape, the proposed approach reduces sensitivity to initialization and improves the robustness and efficiency of HD-DMFT calculations. Moreover, beyond benchmarks within the generated database, we test the model on the actual target problem of bath fitting for Sr$_2$RuO$_4$ (SRO) in a fully interacting DMFT calculation solved with an adaptive-truncation impurity solver, and observe consistently superior performance compared to heuristic initialization.

The remainder of this paper is organized as follows.
Section~\ref{sec:background} introduces the theoretical framework of HD-DMFT and formulates bath fitting as a nonlinear optimization task.
Section~\ref{sec:ml_approach} describes the data-driven initialization methodology, including dataset construction, feature engineering, and the machine-learning model.
Section~\ref{sec:results} presents numerical results demonstrating the effectiveness of the proposed approach.
Finally, Section~\ref{sec:conclusion} summarizes the main findings and discusses future directions.

\section{Background}
\label{sec:background}

\begin{figure*}[h]
\centering
\includegraphics[width=\textwidth]{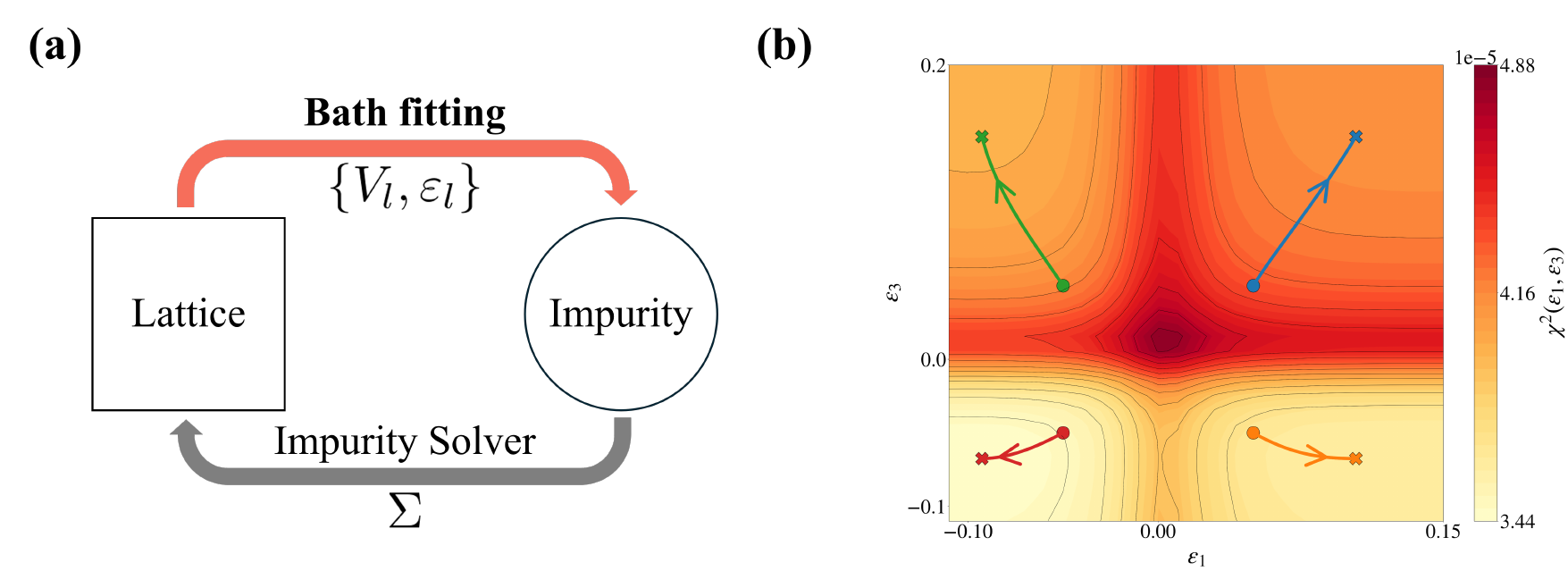}
\caption{
  (a) Schematic overview of the DMFT self-consistency cycle, highlighting the two computational challenges: the impurity solver and bath fitting. Bath fitting constitutes a highly non-convex optimization problem whose convergence behavior depends critically on the quality of the initial bath parameters.
  (b) Cost function $\chi^2$ landscape as a function of two bath onsite energies $(\varepsilon_1, \varepsilon_3)$. Filled circles denote initial parameter values and crosses mark the converged solutions reached by conjugate gradient optimization. Despite starting from nearby positions in parameter space, different trajectories converge to distinct local minima with substantially different cost function values, illustrating the sensitivity of bath fitting to initialization and motivating the need for data-driven strategies to identify favorable starting points.
}
\label{fig:motivation}
\end{figure*}

\subsection{DMFT formalism}
\label{subsec:dmft_overview}

DMFT provides a non-perturbative framework for treating local electronic correlations in lattice models of interacting electrons~\citep{Georges1996,Kotliar2006}. The central approximation of (single-site) DMFT is that the electronic self-energy $\hat{\Sigma}(\mathbf{k}, i\omega_n)$ is purely local, i.e., momentum-independent:
\begin{equation}
  \hat{\Sigma}(\mathbf{k}, i\omega_n) \approx \hat{\Sigma}_{\mathrm{loc}}(i\omega_n),
  \label{eq:local_self_energy}
\end{equation}
where $i\omega_n = i(2n+1)\pi/\beta$ denotes the fermionic Matsubara frequencies and $\beta$ is the fictitious inverse temperature. 
While this locality assumption can be systematically relaxed in cluster extensions of DMFT~\citep{FlorezAblan2025}, we focus throughout this work on the single-site formulation.

Under this approximation, the lattice problem maps onto an auxiliary quantum impurity model consisting of a single correlated site embedded in a self-consistently determined effective medium. The key quantity encoding the coupling between the impurity and its environment is the hybridization function $\hat{\Delta}(i\omega_n)$. Here and throughout, a hat $\hat{(\cdot)}$ denotes matrices in orbital space (e.g., $\hat{G}$, $\hat{\Sigma}$, and $\hat{\Delta}$); in single-orbital systems these objects reduce to scalars. Importantly, the continuous hybridization function encodes the lattice Hamiltonian information through the DMFT self-consistency condition, since $\hat{G}_{\mathrm{loc}}(i\omega_n)$ entering Eq.~\eqref{eq:local_green} is obtained by Brillouin-zone integration of the dressed lattice propagator. The hybridization function is then determined from the updated Weiss field via the Dyson relation in Eq.~\eqref{eq:weiss_update}.

As illustrated in Fig.~\ref{fig:motivation}(a), the DMFT calculation proceeds through an iterative self-consistency cycle between two computational challenges: the impurity solver and bath fitting. 
Starting from an initial set of bath parameters $\pmb{\theta} = \{\varepsilon_l, V_{\mu l}\}_{l=1,\dots,N_{\mathrm{b}};\, \mu=1,\dots,N_{\mathrm{orb}}}$ --- the collection of bath onsite energies $\varepsilon_l$ and hybridization amplitudes $V_{\mu l}$ that together parameterize the discrete bath of the impurity model --- the cycle iterates until convergence. 
The impurity solver---typically based on Hilbert-space diagonalization methods such as exact diagonalization (ED) or configuration interaction (CI)---computes the many-body impurity Green's function $\hat{G}_{\mathrm{imp}}(i\omega_n)$ and self-energy $\hat{\Sigma}_{\mathrm{imp}}(i\omega_n)$ for the given bath configuration. This impurity self-energy is identified with the local lattice self-energy according to Eq.~\eqref{eq:local_self_energy}, and the local lattice Green's function is computed by integrating the dressed lattice propagator over the Brillouin zone:

\begin{equation}
  \hat{G}_{\mathrm{loc}}(i\omega_n) = \int_{\mathrm{BZ}} \frac{d\mathbf{k}}{{V}_{\mathrm{BZ}}} \, \frac{1}{\left[ (i\omega_n + E_{\mathrm{F}})\hat{I} - \hat{H}_0(\mathbf{k}) - \hat{\Sigma}_{\mathrm{imp}}(i\omega_n) \right]},
  \label{eq:local_green}
\end{equation}
where ${V}_{\mathrm{BZ}}$ denotes the Brillouin-zone volume, $E_{\mathrm{F}}$ is the Fermi energy, $\hat{I}$ is the identity matrix in orbital space, and $\hat{H}_0(\mathbf{k})$ is the non-interacting lattice Hamiltonian. The updated Weiss field $\hat{\mathcal{G}}_{0}(i\omega_n)$, which serves as the non-interacting Green's function of the impurity problem, is then obtained via the Dyson equation:
\begin{equation}  \hat{\mathcal{G}}_{0}^{-1}(i\omega_n) = \hat{G}_{\mathrm{loc}}^{-1}(i\omega_n) + \hat{\Sigma}^{}_{\mathrm{imp}}(i\omega_n).
  \label{eq:weiss_update}
\end{equation}
The second computational challenge, bath fitting, determines a new set of bath parameters $\pmb{\theta}$ by minimizing the discrepancy between the discrete bath representation and the continuous hybridization function derived from the updated Weiss field. This optimization is performed at every DMFT iteration and constitutes a highly non-convex problem whose convergence behavior depends sensitively on the quality of the initial bath parameters. If the self-consistency criterion---typically requiring that the change in self-energy or Green's function between successive iterations falls below a prescribed tolerance---is not satisfied, the cycle repeats with the newly optimized bath parameters until convergence is achieved.

The interplay between these two components is central to the computational efficiency of HD-DMFT. While the impurity solver provides the many-body solution for a given bath configuration and represents the primary computational expense for strongly correlated systems, bath fitting provides the numerical bridge from the continuous lattice hybridization function to the discrete bath representation required by the solver. In practice, bath fitting is performed at every DMFT iteration and can dominate the total computational cost when the optimization converges slowly or requires manual intervention due to poor initialization. This observation motivates the development of systematic initialization strategies that improve the robustness and efficiency of the bath-fitting procedure.

\subsection{Bath fitting as a computational bottleneck}
\label{subsec:bath_fitting_bottleneck}
Despite its conceptual simplicity---matching a discrete hybridization function to a continuous target---bath fitting presents formidable computational challenges that can severely limit the practical applicability of HD-DMFT. It is crucial to emphasize, however, that the discrete-bath expression in Eq.~\eqref{eq:discrete_hyb} is not merely used to ``invert'' $\hat{\Delta}(i\omega_n)$ in a generic sense. Rather, it is the concrete representation of the DMFT mapping from the original lattice problem to an auxiliary impurity problem with a finite set of bath parameters: one seeks bath parameters whose discrete hybridization reproduces the lattice-derived continuous $\hat{\Delta}(i\omega_n)$, thereby preserving the lattice information embedded by Eq.~\eqref{eq:local_green} within the impurity model.
The core difficulty arises from the inherently non-convex nature of the optimization landscape.

Fig.~\ref{fig:motivation}(b) illustrates this challenge by visualizing the cost function $\chi^2$, defined formally in Eq.~\eqref{eq:cost_function}, as a function of only two bath onsite energies $(\varepsilon_1, \varepsilon_3)$, with all other bath parameters fixed at their converged values.
Even in this reduced two-dimensional subspace, the landscape exhibits multiple local minima separated by substantial barriers.

The filled circles in Fig.~\ref{fig:motivation}(b) represent different initial parameter values, while the crosses mark the corresponding solutions obtained after conjugate gradient optimization.
Optimization trajectories starting from nearby initial conditions converge to distinct local minima with substantially different cost function values.
This sensitivity demonstrates that in the full high-dimensional parameter space---which can contain dozens or even hundreds of optimization variables for realistic multi-orbital systems---the risk of entrapment in suboptimal local minima is severe.

The consequences of this non-convexity are twofold.
First, unfavorable initial guesses can lead to slow convergence or complete stagnation, requiring manual intervention or computationally expensive multi-start strategies to escape poor local minima.
Second, even when convergence is achieved, there is no guarantee that the solution represents a physically meaningful configuration rather than an artifact of the optimization procedure.
These difficulties are compounded by the fact that bath fitting must be performed repeatedly within the DMFT self-consistency loop, amplifying any inefficiencies in the optimization procedure.
To make these issues concrete, the remainder of this section formulates the finite-bath impurity model used in HD-based solvers and provides the explicit bath-fitting objective optimized in practice.

\subsection{The Anderson impurity model and finite-bath representation}
\label{subsec:aim}

Within HD-DMFT, the quantum impurity problem takes the form of an Anderson impurity model (AIM) with a finite number of bath orbitals~\citep{Caffarel1994,Liebsch2011}. The AIM Hamiltonian consists of three terms:
\begin{equation}
  {H}_{\mathrm{AIM}} = {H}_{\mathrm{imp}} + {H}_{\mathrm{bath}} + {H}_{\mathrm{hyb}}.
  \label{eq:aim_hamiltonian}
\end{equation}
The impurity term ${H}_{\mathrm{imp}}$ describes the correlated orbitals:
\begin{equation}
  {H}_{\mathrm{imp}} = \sum_{\mu\nu,\sigma} [\hat{E}_{\mathrm{imp}}]^{}_{\mu\nu} \, d_{\mu\sigma}^\dagger d^{}_{\nu\sigma} + {H}^{}_{\mathrm{int}},
  \label{eq:h_imp}
\end{equation}
where $d_{\mu\sigma}^\dagger$ ($d_{\mu\sigma}$) creates (annihilates) an electron with spin $\sigma$ in impurity orbital $\mu$, $\hat{E}_{\mathrm{imp}}$ is the on-site impurity orbital energy matrix, and ${H}_{\mathrm{int}}$ contains the local Coulomb interactions~\citep{Georges2013,Werner2006}. 
The bath term ${H}_{\mathrm{bath}}$ represents non-interacting auxiliary orbitals:
\begin{equation}
  {H}_{\mathrm{bath}} = \sum_{l,\sigma} \varepsilon^{}_l \, a_{l\sigma}^\dagger a^{}_{l\sigma},
  \label{eq:h_bath}
\end{equation}
where $a_{l\sigma}^\dagger$ ($a_{l\sigma}$) creates (annihilates) an electron in bath orbital $l$ with energy $\varepsilon_l$.
The hybridization term ${H}_{\mathrm{hyb}}$ couples impurity and bath orbitals:
\begin{equation}
  {H}_{\mathrm{hyb}} = \sum_{\mu,l,\sigma} \left(V^{}_{\mu l} \, d_{\mu\sigma}^\dagger a^{}_{l\sigma} + V_{\mu l}^{*} \, a_{l\sigma}^\dagger d^{}_{\mu\sigma} \right),  \label{eq:h_hyb}
\end{equation}
where $V_{\mu l}$ is an element of the hybridization-strength matrix connecting impurity orbital $\mu$ to bath orbital $l$.

The Weiss field ${\hat{\mathcal{G}}}_0(i\omega_n)$ is related to the hybridization function by:
\begin{equation}
  {\hat{\mathcal{G}}}_0^{-1}(i\omega_n)
  =
  (i\omega_n + E_{\mathrm{F}}) \hat{I}
  - \hat{E}_{\mathrm{imp}}
  - \hat{\Delta}(i\omega_n)
  \label{eq:weiss_field}
\end{equation}
For a finite bath with $N_{\mathrm{b}}$ bath sites, the hybridization function takes the discrete form:
\begin{equation}
  \hat{\Delta}_{\mathrm{bath}}(i\omega_n)
  =
  \sum_{l=1}^{N_{\mathrm{b}}} \frac{\hat{V}^{}_{l} \hat{V}_{l}^{\dagger}}{i\omega_n - \varepsilon_l}.
  \label{eq:discrete_hyb}
\end{equation}
This expression reveals the pole structure of the finite-bath hybridization: each bath orbital contributes a single pole at energy $\varepsilon_l$ with residue matrix $\hat{V}_l \hat{V}_l^\dagger$, where $\hat{V}_l$ is the $l$th bath-coupling column vector.

The accuracy of the discrete representation improves as $N_{\mathrm{b}}$ increases, since additional poles can more faithfully reproduce the frequency structure of the continuous target hybridization function.
However, the Hilbert space dimension of the AIM grows exponentially as $\mathcal{O}(4^{N_{\mathrm{orb}} + N_{\mathrm{b}}})$ for a system with $N_{\mathrm{orb}}$ impurity orbitals and $N_{\mathrm{b}}$ bath sites, placing severe constraints on the maximum bath size tractable with full ED~\citep{Liebsch2011,Amaricci2022,Lorenzo2025}.
In DMFT calculations employing ED solvers, $N_{\mathrm{b}}$ is therefore restricted in practice, and the computational burden shifts from merely increasing bath size to identifying near-optimal bath parameters within a fixed, small bath representation~\citep{FlorezAblan2025}. 
In contrast, configuration-interaction solvers and other truncated-diagonalization variants can accommodate substantially larger bath representations by working in reduced many-body spaces~\citep{Zgid2012,Lin2013}, which increases the dimension of the bath-fitting problem and further amplifies the importance of robust, data-driven initialization.

\subsection{Formulation and challenges of the bath-fitting problem}
\label{subsec:bath_fitting}

The bath-fitting procedure provides a numerical bridge from the continuous hybridization function ${\hat{\Delta}}_{\mathrm{latt}}(i\omega_n)$ derived from the DMFT self-consistency condition to the finite-bath approximation ${\hat{\Delta}}_{\mathrm{bath}}(i\omega_n;\pmb{\theta})$ required by HD-based impurity solvers. The lattice hybridization function is extracted from the updated Weiss field via
\begin{equation}
\hat{\Delta}_{\mathrm{latt}}(i\omega_n) = (i\omega_n + E_{\mathrm{F}})\hat{I} - \hat{E}_{\mathrm{imp}} - \hat{\mathcal{G}}_{0,\mathrm{new}}^{-1}(i\omega_n).
  \label{eq:lattice_hyb}
\end{equation}
The bath fitting problem then seeks the set of bath parameters $\pmb{\theta}$ that minimizes the cost function
\begin{equation}
  \chi^2(\pmb{\theta}) = \frac{1}{N_\omega}\sum_{n=0}^{N_\omega - 1} 
  \left\| 
  \hat{\mathcal{G}}_{0,\mathrm{new}}^{-1}(i\omega_n) 
  - 
  \hat{\mathcal{G}}_{0,\mathrm{bath}}^{-1}(i\omega_n; \pmb{\theta}) 
  \right\|_{\mathrm{F}}^2,
  \label{eq:cost_function}
\end{equation}
where $N_\omega$ denotes the number of Matsubara frequencies included in the fit and $\|\cdot\|_{\mathrm{F}}$ is the Frobenius norm. The discrete inverse Weiss field appearing in Eq.~\eqref{eq:cost_function} is constructed from the bath parameters as
\begin{equation}
  \hat{\mathcal{G}}_{0,\mathrm{bath}}^{-1}(i\omega_n; \pmb{\theta})
  =
  (i\omega_n + E_{\mathrm{F}})\hat{I}
  -
  \hat{E}_{\mathrm{imp}}
  -
  \hat{\Delta}_{\mathrm{bath}}(i\omega_n; \pmb{\theta}).
  \label{eq:discrete_weiss_inv}
\end{equation}
Equivalently, the cost function may be expressed directly in terms of the hybridization functions as
\begin{equation}
  \tilde{\chi}^2(\pmb{\theta}) = 
  \frac{1}{N_\omega}\sum_{n=0}^{N_\omega - 1} 
  \left\| \hat{R}(i\omega_n,\pmb{\theta}) \right\|_{\mathrm{F}}^{2},
  \label{eq:cost_function_hyb}
\end{equation}
where $\hat{R}(i\omega_n,\pmb{\theta}) = \hat{\Delta}_{\mathrm{latt}}(i\omega_n) - \hat{\Delta}_{\mathrm{bath}}(i\omega_n, \pmb{\theta})$ is the residual matrix quantifying the mismatch between the lattice and discrete hybridization functions.
These two formulations differ only by a constant term and therefore share identical gradients with respect to the bath parameters. In the present work, the conjugate gradient optimization employs the Weiss-field formulation in Eq.~\eqref{eq:cost_function}, while $\tilde{\chi}^2$ defined in Eq.~\eqref{eq:cost_function_hyb} is reported when assessing initialization quality and convergence performance, as it isolates the hybridization fitting error.

The total number of fitting parameters depends on the bath structure and the imposed symmetry constraints. 
For a general multi-orbital system with $N_{\mathrm{orb}}$ impurity orbitals and $N_{\mathrm{b}}$ bath sites, where each bath site couples to all impurity orbitals, the parameter count is
\begin{equation}
  N_{\mathrm{params}} = 2 N_{\mathrm{b}} (4 N_{\mathrm{orb}} + 1),
  \label{eq:num_params}
\end{equation}
which comprises $2N_{\mathrm{b}}$ real bath onsite energies $\{\varepsilon_l\}$ (accounting for two spin channels) and $2N_{\mathrm{b}} \times 2N_{\mathrm{orb}}$ complex hybridization strengths $\{V_{\mu l}\}$ (connecting bath and impurity spin-orbitals). Each complex hybridization strength contributes two independent real parameters, which can be represented either in Cartesian form (real and imaginary parts) or in polar form (magnitude and phase angle), as discussed in Section~\ref{sec:ml_approach}. This count assumes no symmetry constraints; physical symmetries such as time-reversal invariance can substantially reduce the effective dimensionality, as discussed in Section~\ref{sec:ml_approach}.

The optimization problem defined by Eq.~\eqref{eq:cost_function} is inherently nonlinear and non-convex due to the functional form of the discrete hybridization function. The bath onsite energies $\varepsilon_l$ enter through the denominators in Eq.~\eqref{eq:discrete_hyb}, while the hybridization strengths $V_{\mu l}$ appear in the numerators. Consequently, the cost function consists of a sum of squared residuals involving rational functions of the bath parameters, rendering the optimization landscape rugged and typically populated by multiple local minima. This structure, already visualized in Fig.~\ref{fig:motivation}(b) for a minimal two-dimensional slice, explains the strong sensitivity of the converged solution to the choice of initial bath parameters.

Several structural features contribute to the difficulty of the bath fitting problem. The presence of terms of the form $(i\omega_n - \varepsilon_l)^{-1}$ makes the objective function non-quadratic, and gradient-based optimization methods can easily become trapped in initialization-dependent local minima without any guarantee of reaching the global optimum~\citep{MejutoZaera2020,Liebsch2011}. The high dimensionality of the parameter space---which can exceed 100 variables for realistic multi-orbital systems---further exacerbates the optimization challenge. As a result, bath fitting is widely regarded as one of the most delicate steps in the HD-DMFT workflow, frequently requiring expert intuition to achieve reliable convergence~\citep{Senechal2010}.

In addition to the intrinsic nonlinearity of the cost function, physical symmetries impose important constraints on the bath parameterization. These symmetries reduce the effective dimensionality of the optimization problem and improve numerical stability~\citep{Koch2008,Senechal2010}. 
In the present work, time-reversal symmetry is incorporated, as it is relevant for the non-magnetic layered ruthenate models under consideration. 
This symmetry implies that bath onsite energies and hybridization strengths occur in time-reversed pairs with identical magnitudes, and that the global phase of each hybridization strength $V_{\mu l}$ is physically irrelevant. 
These constraints are explicitly enforced within the machine learning framework through symmetry-aware feature construction and output normalization, as discussed in Section~\ref{sec:ml_approach}.

Despite the complexity of the optimization landscape, the gradient of the cost function with respect to the bath parameters can be computed analytically. 
For a given bath onsite energy $\varepsilon_m$, the derivative of the hybridization function takes the form
\begin{equation}
  \frac{\partial {\big[\hat{\Delta}_{\mathrm{bath}}(i\omega_n,\pmb{\theta})\big]}_{\mu\nu}}{\partial \varepsilon_m}
    =
  \frac{V^{}_{\mu m} V_{\nu m}^{*}}{(i\omega_n - \varepsilon_m)^2},
  \label{eq:d_delta_d_eps}
\end{equation}
which yields the corresponding gradient component
\begin{equation}
  \frac{\partial \tilde{\chi}^2}{\partial \varepsilon_m}
  =
  \frac{2}{N_\omega}
  \sum_{n=0}^{N_\omega - 1} 
  \sum_{\mu,\nu}
  \mathrm{Re}
  \left[
    \hat{R}^{\dagger}_{\mu\nu}(i\omega_n)
  \frac{V^{}_{\mu m} V_{\nu m}^{*}}{(i\omega_n - \varepsilon_m)^2}
  \right].
  \label{eq:grad_eps}
\end{equation}
The derivatives with respect to the real and imaginary parts of the complex hybridization strengths $V_{\mu l}$ follow analogously by application of the chain rule. The availability of analytical gradients enables the use of efficient gradient-based optimization methods and avoids the numerical inaccuracies associated with finite-difference schemes.

Starting from an initial guess $\pmb{\theta}^{(0)}$, the bath parameters are optimized by minimizing the cost function using the Fletcher--Reeves nonlinear conjugate gradient algorithm~\citep{Fletcher1964}. This method generalizes the linear conjugate gradient approach to smooth nonlinear objective functions and exhibits favorable convergence properties without requiring explicit evaluation or storage of the Hessian matrix. The algorithm generates a sequence of iterates according to
\begin{equation}
  \pmb{\theta}^{(k+1)} =
  \pmb{\theta}^{(k)} +
  \alpha_k \mathbf{d}^{(k)},
  \label{eq:cg_update}
\end{equation}
where $\alpha_k > 0$ is a step size determined by a line search and $\mathbf{d}^{(k)}$ denotes the search direction. The initial direction is chosen as the steepest descent direction, $\mathbf{d}^{(0)} = -\nabla \chi^2(\pmb{\theta}^{(0)})$, and subsequent directions are updated according to
\begin{equation}
  \mathbf{d}^{(k)} =
  -\nabla \chi^2(\pmb{\theta}^{(k)}) +
  \beta_k^{\mathrm{FR}} \mathbf{d}^{(k-1)},
  \label{eq:cg_direction}
\end{equation}
with the Fletcher--Reeves coefficient
\begin{equation}
  \beta_k^{\mathrm{FR}} =
  \frac{\big|\nabla \chi^2(\pmb{\theta}^{(k)})\big|^2}
  {\big|\nabla \chi^2(\pmb{\theta}^{(k-1)})\big|^2}.
  \label{eq:fletcher_reeves}
\end{equation}
In practice, the step size $\alpha_k$ is determined by a line search satisfying the Wolfe conditions. Convergence is declared when either the relative change in the cost function between successive iterations falls below a prescribed tolerance, ${\big|\chi^2\big(\pmb{\theta}^{(k)}\big) - \chi^2\big(\pmb{\theta}^{(k-1)}\big)\big|}/{\chi^2\big(\pmb{\theta}^{(k)}\big)} < 10^{-5}$, or the gradient norm becomes sufficiently small, $\big|\nabla \chi^2\big(\pmb{\theta}^{(k)}\big)\big| < \epsilon_g = 10^{-8}$.

Because bath fitting is performed at every iteration of the DMFT self-consistency loop, computational efficiency is a critical consideration. An unfavorable initialization may lead the optimization into a suboptimal local minimum, necessitating manual intervention or computationally expensive multi-start strategies. In contrast, an initial guess that lies within the basin of attraction of a physically relevant minimum can substantially reduce the number of optimization steps and improve the robustness of the overall DMFT calculation. These considerations motivate the data-driven initialization strategy introduced in Section~\ref{sec:ml_approach}, which aims to provide high-quality initial guesses that:
\begin{itemize}
\item Reduce the initial fitting error $\tilde{\chi}^2_\mathrm{init}\big(\pmb{\theta}^{(0)}\big)$ compared to heuristic initialization
\item Decrease the number of conjugate gradient iterations $N_{\mathrm{iter}}$ required for convergence
\item Maintain or improve the quality of the final converged solution $\tilde{\chi}^2_\mathrm{final}\big(\pmb{\theta}^{\mathrm{final}}\big)$
\end{itemize}
By learning the complex high-dimensional mapping from hybridization functions to near-optimal bath parameters through supervised machine learning, we aim to systematically avoid the pitfalls illustrated in Fig.~\ref{fig:motivation}(b) and provide a robust, automated alternative to expert-guided manual initialization.

\section{Machine learning approach}
\label{sec:ml_approach}

The bath fitting problem introduced in Section~\ref{subsec:bath_fitting} can be reformulated as a supervised learning task. The central idea is to train a regression model that maps the continuous hybridization function to an initial set of bath parameters, thereby providing a physically informed starting point for the subsequent nonlinear optimization.

In our earlier proof-of-concept study for single-orbital systems~\citep{Kim2024}, we demonstrated that the success of such machine learning initialization depends critically on the structure of the training data: models trained on bath parameters with regular energy distributions achieved high predictive accuracy, whereas randomly distributed bath energies led to poor generalization regardless of training set size. This observation motivates the physically grounded data generation strategy developed in Section~\ref{subsec:data_generation}, where training labels are obtained from fully converged optimizations rather than random sampling.

This section describes the data-driven initialization framework developed in this work and introduces a heuristic initialization method that serves as a baseline for performance comparison. All KRR-models employ a radial basis function (RBF) kernel with regularization parameter $\alpha = 10^{-2}$ and length scale $\gamma = 1/n^{}_{\mathrm{features}}$, where $n^{}_{\mathrm{features}}$ denotes the input feature dimension. The dataset is partitioned using an $8{:}2$ train--test split.

\subsection{Regression formulation}
\label{subsec:regression_formulation}

\subsubsection{Learning objective}
\label{subsubsec:learning_objective}

The objective of the supervised learning task is to predict bath parameters that serve as high-quality initial conditions for the conjugate gradient optimization. 
Let $\pmb{\theta}_X^{(0)}$ denote an initial set of bath parameters obtained from a chosen initialization scheme $X$ for a given lattice hybridization function. 
The quality of this initialization is assessed through three complementary metrics.

First, the initial cost function value $\tilde{\chi}^2_{\text{init}}$ evaluated at the predicted parameters quantifies how close the ML prediction lies to the optimal solution in parameter space. 
A lower initial cost function indicates that the optimization begins within the basin of attraction of a physically relevant local minimum.

Second, the number of conjugate gradient iterations $N_{\text{iter}}$ required to reach convergence measures the computational savings achieved by the improved initialization. 
The iteration count directly impacts the computational time of the bath fitting step within each DMFT self-consistency cycle.

Third, the final cost function $\tilde{\chi}^2_{\text{final}}$ after optimization verifies that the converged solution attains comparable or better accuracy than that obtained from heuristic initialization. 
This metric ensures that the reduced iteration count reflects genuine computational savings enabled by an automated, physically informed bath initialization, rather than premature termination at a suboptimal local minimum.

The ML initialization is considered successful if it demonstrates superior performance across all three metrics: lower initial cost function values, reduced iteration counts to convergence, and comparable or improved final converged solutions relative to the baseline heuristic initialization method described in Section~\ref{subsec:heuristic}. 
This multi-metric evaluation framework ensures that the data-driven approach provides both computational efficiency and solution quality.

\subsubsection{Time-reversal symmetry constraints}
\label{sec:trs_constraints}

A key feature of our approach is the systematic incorporation of time-reversal symmetry, which reduces the effective dimensionality of both the input and output spaces by nearly a factor of two.
For $2N_\mathrm{b}$ bath spin-orbitals, we label bath Kramers pairs as
$(2l-1,2l)$ with $l=1,\ldots,N_\mathrm{b}$.
Time-reversal symmetry enforces Kramers degeneracy of the bath onsite energies,
\begin{equation}
  \varepsilon^{}_{2l-1} = \varepsilon^{}_{2l},
  \qquad l = 1,\ldots, N_\mathrm{b}.
  \label{eq:trs_bath_energy}
\end{equation}
Writing the hybridization strengths in polar form,
$\hat{V}_{\mu l} = |{V}_{\mu l}| e^{i\theta_{\mu l}}$,
it is instructive to impose time-reversal symmetry directly at the level of the microscopic
hybridization parameters.
In a general bath basis, the action of the time-reversal operator within each
bath Kramers doublet is defined up to a pair-dependent $U(1)$ gauge phase.
Accordingly, time-reversal invariance of the hybridization Hamiltonian implies
\begin{equation}
  {V}^{}_{\bar{\mu},\,2l}
  =
  \eta^{}_\mu\, e^{-i\gamma^{}_l}\,
  {V}^{\dagger}_{\mu,\,2l-1},
  \qquad
  l = 1,\ldots,N_\mathrm{b},
  \label{eq:trs_V_basic}
\end{equation}
where $\bar{\mu}$ denotes the Kramers partner of orbital $\mu$,
$\eta_\mu \in \{+1,-1\}$ is the time-reversal phase factor determined by the
$j$-effective impurity basis, and $\gamma_l\in[0,2\pi)$ is a bath-pair-dependent
gauge phase.
The phases $\gamma_l$ can always be absorbed by a rephasing of the bath operators
within each Kramers doublet; we keep them explicit here because they clarify how
time-reversal symmetry constraints appear in generic numerical parameterizations.
Taking the modulus of Eq.~\eqref{eq:trs_V_basic} immediately yields the magnitude constraint
\begin{equation}
  |{V}^{}_{\bar{\mu},\,2l}| = |{V}^{}_{\mu,\,2l-1}|,
  \qquad l = 1, \ldots, N_{\mathrm{b}}.
  \label{eq:trs_hyb_magnitude}
\end{equation}
Substituting the polar form into Eq.~\eqref{eq:trs_V_basic} and comparing phases gives
\begin{equation}
  \theta^{}_{\bar{\mu},\,2l}
  = -\theta^{}_{\mu,\,2l-1} + \varphi^{}_\mu - \gamma^{}_l\quad
  \pmod{2\pi},
  \label{eq:trs_phase_single}
\end{equation}
where $\eta_\mu = e^{i\varphi_\mu}$ with $\varphi_\mu \in \{0,\pi\}$.
Subtracting Eq.~\eqref{eq:trs_phase_single} for two orbitals $\mu$ and $\nu$
eliminates the bath-pair phase $\gamma_l$ and yields the gauge-invariant relative-phase constraint
\begin{equation}
  \theta^{}_{\mu,\,2l-1} - \theta^{}_{\nu,\,2l-1}
  = -\bigl(\theta^{}_{\bar{\mu},\,2l} - \theta^{}_{\bar{\nu},\,2l}\bigr) + \phi^{}_{\mu\nu}\quad
  \pmod{2\pi},
  \label{eq:trs_hyb_phase}
\end{equation}
where
\begin{equation}
  \phi^{}_{\mu\nu}
  \equiv \varphi^{}_\mu - \varphi^{}_\nu
  = \frac{\pi(1 - \eta^{}_\mu \eta^{}_\nu)}{2}
  \in \{0, \pi\}.
  \label{eq:trs_phi}
\end{equation}

Eqs.~\eqref{eq:trs_V_basic}--\eqref{eq:trs_phi} establish that once the bath parameters
for one representative of each Kramers pair are specified, the partner amplitudes are fixed by time-reversal symmetry.
Importantly, any overall phase ambiguity associated with a bath Kramers pair cancels in
Eq.~\eqref{eq:trs_hyb_phase}, rendering it directly verifiable and suitable for symmetry-aware
feature/target construction.

\subsubsection{Input and output representation}
\label{sec:inout_representation}

The $t_{2g}\otimes\mathrm{spin}$ impurity Hilbert space is represented in a $j$-effective basis only as a unitary basis choice. No explicit spin--orbit coupling is included in either the tight-binding Hamiltonian or the bath-fitting calculations. This basis is used to make the Kramers-pair structure and time-reversal constraints transparent, while remaining fully equivalent to the bare orbital-spin basis.

\begin{figure*}[!htbp]
\centering
\includegraphics[width=\textwidth]{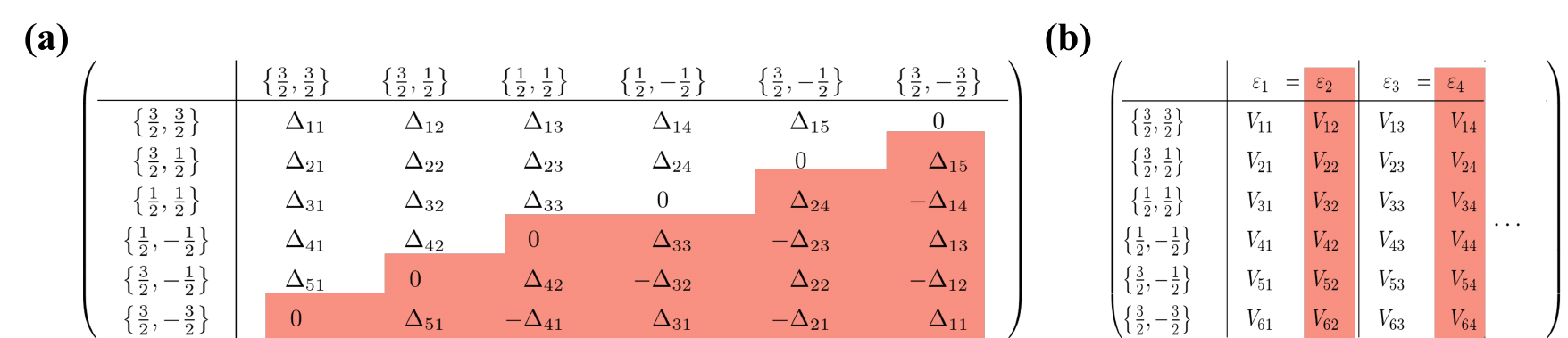}
\caption{
    Schematic representation of (a) the hybridization-function matrix and (b) the bath-parameter matrix.
    Only the symmetry-independent (unshaded) components are used as machine-learning inputs and outputs.
    The remaining (shaded) elements are related by time-reversal (Kramers) symmetry and are reconstructed a posteriori,
    rather than learned independently.
}
\label{fig:data_rep}
\end{figure*}

The input to the machine learning model is the lattice hybridization function $\hat{\Delta}_{\text{latt}}(i\omega_n)$ 
evaluated on the first $N_\omega$ fermionic Matsubara frequencies within the fitting window. Without imposing 
symmetry constraints, each frequency contributes a $6 \times 6$ complex matrix, corresponding to 72 
independent real numbers per frequency when both real and imaginary parts are counted.

Time-reversal symmetry substantially reduces this dimensionality. 
Fig.~\ref{fig:data_rep}(a) illustrates the resulting partitioning of the hybridization-function matrix: 
the unshaded region contains the symmetry-independent complex elements retained as input features, 
whereas the shaded region consists of dependent entries that are reconstructed from their 
Kramers-paired counterparts through the time-reversal symmetry relations. 
In the present case, separating real and imaginary parts, the input feature vector for a single sample has dimension
\begin{equation}
  d_{\text{feature}} = 4N_\omega(N_\mathrm{orb})^2 
  \label{eq:feature_dim}
\end{equation}
corresponding to a factor-of-two reduction compared to the naive representation, 
as illustrated in Fig.~\ref{fig:data_rep}(a). 
The dataset comprises $N_{\text{samples}} = 1024$ structurally distinct configurations generated 
through the procedure described in Section~\ref{subsec:data_generation}.

The output of the regression model consists of symmetry-reduced bath parameters.
Applying the constraints in Eqs.~\eqref{eq:trs_bath_energy}--\eqref{eq:trs_hyb_phase},
the independent parameters are those of one representative in each bath Kramers pair.
Concretely, we treat $\{\varepsilon^{}_{2l-1}\}$ and
$\{\hat{V}_{\mu,\,2l-1}\}_{\mu,\,l}$ as independent, and reconstruct the partner bath orbitals
$(2l)$ using Eq.~\eqref{eq:trs_V_basic}.
As shown in Fig.~\ref{fig:data_rep}(b),
the shaded columns are fixed by time-reversal symmetry from their Kramers-paired counterparts. 
Hence, only the remaining (unshaded) symmetry-independent parameters need to be learned.

Fig.~\ref{fig:data_rep}(b) depicts the structure of the bath-parameter matrix, 
where columns correspond to bath orbitals and rows to impurity orbitals. 
Time-reversal symmetry imposes Kramers degeneracy on the bath onsite energies 
Eq.~\eqref{eq:trs_bath_energy}, enforcing $\varepsilon^{}_1=\varepsilon^{}_2$, 
$\varepsilon^{}_3=\varepsilon^{}_4$, and so forth. 
Likewise, the hybridization parameters in the shaded columns are not independent 
but are reconstructed from their Kramers-paired counterparts through 
Eqs.~\eqref{eq:trs_hyb_magnitude}--\eqref{eq:trs_hyb_phase}. 
Consequently, only the remaining (unshaded) symmetry-independent elements---namely 
$2N_\mathrm{b}$ real bath onsite energies and 
$2N_\mathrm{b}\times 2N_{\text{orb}}$ complex hybridization strengths---constitute 
the regression target.

Even after imposing time-reversal symmetry, the mapping from a hybridization function to bath parameters is not unique; we therefore apply the following canonicalization.
First, permuting bath orbital indices leaves $\Delta_{\mathrm{bath}}(i\omega_n)$ invariant because Eq.~\eqref{eq:discrete_hyb} involves a sum over bath sites.
Second, each bath orbital carries an independent $U(1)$ gauge freedom:
multiplying all hybridization strengths $\hat{V}_{\mu l}$ for a given $l$ by a common phase $e^{i\phi_l}$
leaves the product $\hat{V}^{}_{\mu l}\hat{V}_{\nu l}^{\dagger}$ unchanged.

This canonicalization ensures that each hybridization function maps to a 
unique set of bath parameters, rendering the supervised learning problem 
well-posed. The same canonicalization is applied to both training labels 
and model predictions before computing the parity plots in Section~\ref{subsec:prediction_quality}.

Separating amplitude and phase parts for the hybridization strengths, the total number of independent 
output parameters is
\begin{equation}
  d_{\text{label}} = 4N_\mathrm{b}N_\mathrm{orb},
  \label{eq:label_dim}
\end{equation}
which follows from retaining only the symmetry-independent bath parameters and imposing gauge fixing to remove redundant phase degrees of freedom.
During inference, the complete bath parametrization required by the impurity solver is recovered by applying
Eq.~\eqref{eq:trs_bath_energy} together with the time-reversal symmetry reconstruction formula Eq.~\eqref{eq:trs_V_basic}
(and its phase form Eq.~\eqref{eq:trs_phase_single}) to generate the Kramers-partner bath orbitals from the
predicted symmetry-independent parameters.

\subsection{Data generation}
\label{subsec:data_generation}

A central methodological advance of this work compared to our previous study~\citep{Kim2024} is the physically motivated construction of the training dataset. 
Rather than generating bath parameters randomly---an approach that was shown to fail for bath energies without regular structure---we construct training data from non-interacting limit calculations on structurally deformed model ruthenate configurations. 
This ensures that the training labels represent physically realizable bath configurations that have been optimized to accurately represent continuous hybridization functions, rather than arbitrary parameter combinations that may not correspond to any physical situation.
This section describes the tight-binding model, the 
structural sampling procedure, and the generation of input-output pairs for supervised learning.

\subsubsection{Tight-binding model}
\label{sec:tb_model}
The electronic structure of the Ru $t{}_{2g}$ manifold in Ca$_2$RuO$_4$ is
described by a Slater--Koster tight-binding Hamiltonian. We label the
$t^{}_{2g}$ orbitals by the \emph{missing-axis} index $a\in\{x,y,z\}$,
\begin{equation}
d_{\bar x}\equiv d_{yz},\qquad
d_{\bar y}\equiv d_{zx},\qquad
d_{\bar z}\equiv d_{xy},
\end{equation}
and denote spin by $\sigma$. The one-electron lattice Hamiltonian reads
\begin{equation}
{H}_{\mathrm{latt}}
= \sum_{i,\bar a,\sigma} \epsilon^{}_{\bar a}\,
 d^{\dagger}_{i\bar a\sigma} d^{}_{i\bar a\sigma}
+ \sum_{\langle i,i'\rangle}\sum_{\bar a,\bar b,\sigma}
 {T}^{ii'}_{\bar a \bar b}\,
 d^{\dagger}_{i\bar a\sigma} d^{}_{i'\bar b\sigma},
\label{eq:tb_realspace}
\end{equation}
where $\epsilon_{\bar a}$ is the on-site crystal-field energy of orbital $\bar a$, 
and the second term describes inter-site hopping between nearest-neighbor Ru sites.
The total nearest-neighbor hopping matrix $\hat{T}^{ii'}$ comprises two
contributions:
\begin{equation}
\hat{T}^{ii'} = \hat{T}^{ii'}_{dd} + \hat{T}^{ii'}_{dpd},
\label{eq:total_hopping}
\end{equation}
where $\hat{T}^{ii'}_{dd}$ represents direct Ru--Ru $d$--$d$ hopping matrix constructed
from Slater--Koster overlaps $(V_{dd\sigma}, V_{dd\pi}, V_{dd\delta})$, and
$\hat{T}^{ii'}_{dpd}$ represents oxygen-mediated $d$--$p$--$d$ hopping matrix.

The oxygen-mediated contribution
is constructed using matrices $\hat{\tau}^{s}_{j}$ that encode the symmetry-allowed
Ru--O $t^{}_{2g}$--$p$ hopping for an oxygen located at displacement $s\,\hat{j}$
along the local ${j}\in\{x,y,z\}$ axis, with $s=\pm$ reflecting the odd parity
of oxygen $p$ orbitals. All orbital matrix elements below are expressed in the local octahedral axes attached to
each Ru site, whereas the global crystallographic frame enters only through the atomic
coordinates and bond geometry. In the
$t^{}_{2g}$ basis $(d_{\bar x},d_{\bar y},d_{\bar z})$
and oxygen basis $(p_x,p_y,p_z)$, the matrix elements are compactly written as
\begin{equation}
{\big(\hat{\tau}^{s}_{j}\big)}_{a\mu} = s\,V^{}_{pd\pi}\,\varepsilon_{a\mu j}^{\,2},\; a,\mu,j\in\{x,y,z\}, \; \varepsilon_{xyz}=+1,
\label{eq:tau_matrices}
\end{equation}
so that $\varepsilon_{a\mu j}^{2}\in\{0,1\}$ selects the symmetry-allowed
$t^{}_{2g}$--$p$ $\pi$-hopping channels.

\begin{figure*}[h]
\centering
\includegraphics[width=0.95\textwidth]{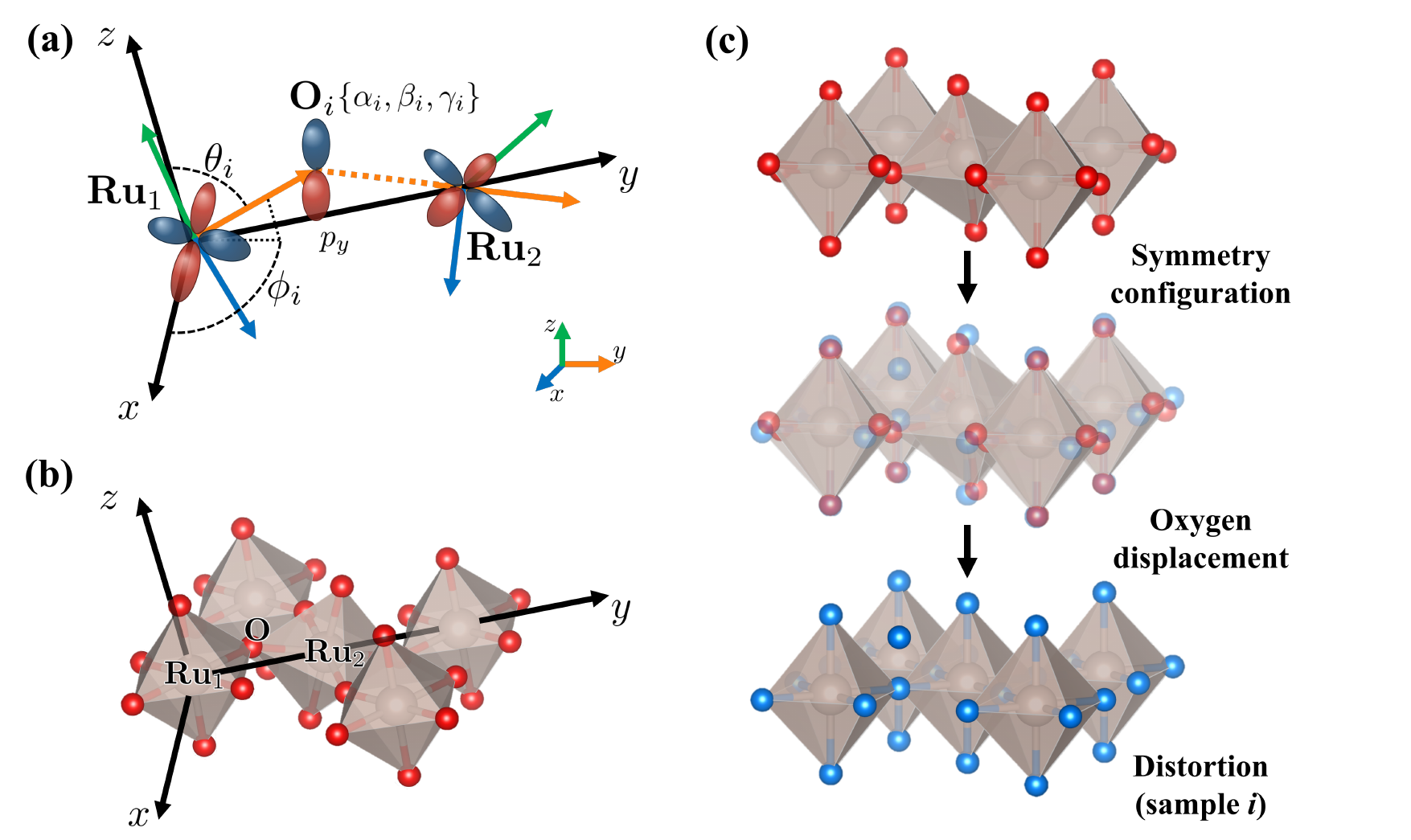}
\caption{%
    Structural sampling for the training database.
    \textbf{(a)} Schematic illustration of the Ru--O--Ru hopping geometry. The oxygen O$_i$ between neighboring Ru sites is controlled by the displacement parameters $(\theta_i)$ and $(\phi_i)$, while the oxygen configuration at the displaced position is also characterized by the Euler angles $({\alpha_i,\beta_i,\gamma_i})$. The coordinate triad shown in the lower-right inset represents the local coordinate frame attached to each Ru site, in which the Ru (d)-orbital basis is defined.    
    \textbf{(b)} Reference layered RuO$_{6}$ network in the global frame.
    \textbf{(c)} Symmetric reference configuration (top, red), overlay of the reference and distorted configurations (middle), and the $i$th distorted configuration (bottom, blue) generated by sampling rotation matrices parameterized by Euler angles. (For interpretation of the references to color in this figure legend, the reader is referred to the web version of this article.)
}
\label{fig:sampling}
\end{figure*}

For a Ru$_i$--O--Ru$_{i'}$ pathway in Fig.~\ref{fig:sampling}(a), the effective hopping is
\begin{equation}
\hat{T}^{ii'}_{dpd}
= \hat{\tau}^{(i\to O)}\,{\big(\hat{R}^{(i)}\big)}^{T} \hat{R}^{(i')}\,{\big(\hat{\tau}^{(i'\to O)}\big)}^{T},
\label{eq:dpd_hop}
\end{equation}
where $\hat{R}^{(i)}$ is the rotation matrix from global to local octahedral axes
at Ru site $i$, so that ${\big(\hat{R}^{(i)}\big)}^{T}\hat{R}^{(i')}$ transforms the
intermediate oxygen $p$-orbital index from the local frame of site $i'$ to that of
site $i$, and $\hat{\tau}^{(i\to O)}\equiv \hat{\tau}^{s_{iO}}_{j_{iO}}$ denotes
the Ru$_i\to$O hopping matrix determined by the local bond direction
$j_{iO}\in\{x,y,z\}$ and sign $s_{iO}=\pm$ (analogously for $i'$).

The oxygen $p$ orbitals are not retained as explicit basis states in
Eq.~\eqref{eq:tb_realspace}. Although Wannier orbitals carrying both Ru $d$
and O $p$ character can reproduce the DFT bands almost exactly, the present
construction targets only the low-energy Ru $t^{}_{2g}$ manifold, into which
the leading effect of the O $p$ states is downfolded as the effective
$d$--$p$--$d$ hopping of Eq.~\eqref{eq:dpd_hop}.

The tight-binding parameters used for dataset generation are listed in 
Table~\ref{tab:tb_params}. Prior to computing the local Green's function 
Eq.~\eqref{eq:weiss_field}, all hopping amplitudes are globally rescaled such 
that the $t^{}_{2g}$ bandwidth satisfies $W = 1$.

\begin{table}[h]
\centering
\caption{Tight-binding parameters for Ca$_2$RuO$_4$, expressed in the local
$t^{}_{2g}$ orbital basis $(yz, zx, xy)$ attached to each RuO$_6$ octahedron.
The Slater--Koster amplitudes $(V_{pd\pi}, V_{dd\sigma}, V_{dd\pi},
V_{dd\delta})$ and the on-site orbital energies $\epsilon_{\bar a}$ are
obtained by fitting to the nearest-neighbor Wannier Hamiltonian; 
The NN hopping matrix elements $t^{\mathrm{NN}}_{\alpha\beta}$ evaluated for the reference structure shown in Fig.~\ref{fig:sampling}(b) are also shown for comparison. All values are in eV prior to bandwidth normalization.}
\label{tab:tb_params}

\centering

\setlength{\tabcolsep}{3pt}
\renewcommand{\arraystretch}{1.15}

\begin{tabularx}{\columnwidth}{@{}*{6}{>{\centering\arraybackslash}X}@{}}
\hline\hline
\multicolumn{2}{c}{Slater--Koster} &
\multicolumn{2}{c}{On-site energies} &
\multicolumn{2}{c}{NN hopping} \\
\hline
$V_{pd\pi}$    & $-0.47$             &
$\epsilon_{yz}$ & $4.81$             &
$t_{yz,zx}^{\mathrm{NN}}$ & $0.01$ \\

$V_{dd\sigma}$ & $\phantom{-}0.40$   &
$\epsilon_{zx}$ & $4.79$             &
$t_{zx,xy}^{\mathrm{NN}}$ & $0.02$ \\

$V_{dd\pi}$    & $\phantom{-}0.11$   &
$\epsilon_{xy}$ & $4.52$             &
$t_{xy,yz}^{\mathrm{NN}}$ & $0.09$ \\

$V_{dd\delta}$ & $-0.04$             &
& &
& \\
\hline\hline
\end{tabularx}
\end{table}

\subsubsection{Structural sampling}
\label{sec:struct_sampling}
The Slater--Koster parameters $(V_{dd\pi}, V_{dd\delta}, V_{pd\pi})$ are determined by fitting the tight-binding band structure to a nearest-neighbor (NN) Wannier Hamiltonian extracted from density functional theory (DFT).
The underlying crystal structure is taken from the neutron diffraction refinement of Ca$_2$RuO$_4$ reported by Braden et al.~\citep{Braden1998}, specifically the low-temperature orthorhombic \textit{Pbca} (S-CRO, 11~K) phase.
The experimental lattice parameters and internal coordinates are adopted directly, without structural relaxation.
For convenience in orbital projection, the orthorhombic in-plane lattice vectors are expressed in a $45^\circ$ rotated representation, which preserves the underlying symmetry while aligning the local axes with the RuO$_6$ octahedral network.

DFT calculations are performed using the Vienna \textit{Ab initio} Simulation Package (VASP)~\citep{Kresse1996,Kresse1999} within the projector augmented-wave framework.
A plane-wave kinetic energy cutoff of 500~eV and an electronic convergence criterion of $10^{-6}$~eV are employed.
Methfessel--Paxton smearing of 0.1~eV is used. The calculations are non-spin-polarized, without spin--orbit coupling or Hubbard-$U$ corrections, and are carried out at fixed experimental geometry.
Maximally localized Wannier functions are constructed using Wannier90~\citep{Mostofi2014}, targeting the Ru $t^{}_{2g}$ subspace without explicitly including oxygen $p$ states.

\begin{figure*}[h]
\centering
\includegraphics[width=\textwidth]{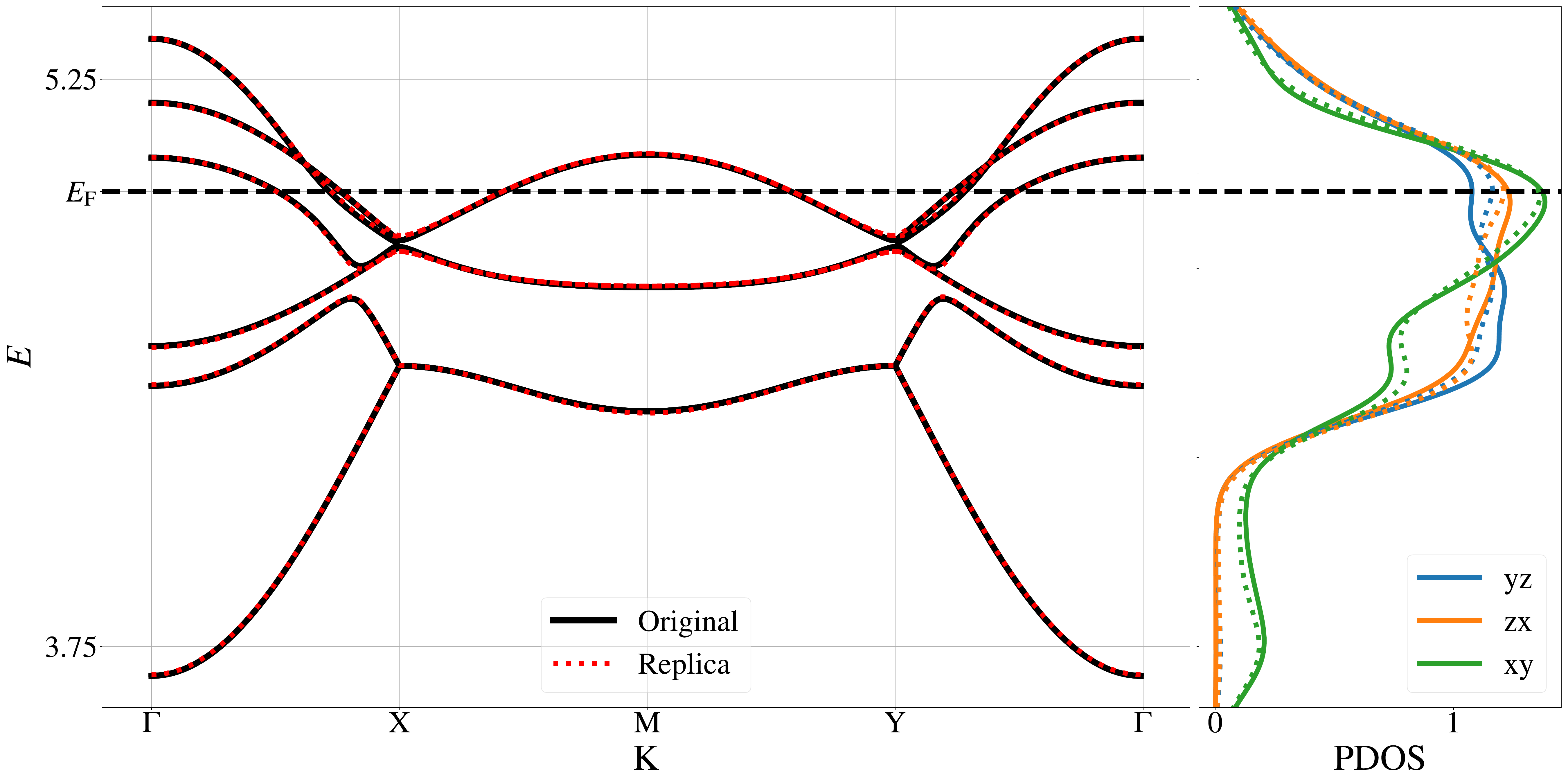}
\caption{%
    Reproduction of the nearest-neighbor DFT band dispersion of
    Ca$_2$RuO$_4$ using fitted Slater--Koster parameters.
    The reference nearest-neighbor hoppings are extracted from the
    Wannier-projected Ru $t_{2g}$ Hamiltonian and then expressed in the
    local $t_{2g}$ basis $(yz,zx,xy)$ attached to the RuO$_6$ octahedra.
    The parameters in Table~\ref{tab:tb_params} are fitted to reproduce
    these local-basis nearest-neighbor hoppings. The comparison is used
    only to obtain physically reasonable hopping amplitudes for dataset
    generation, not to construct a complete Ru-$d$/O-$p$ Wannier model.
}
\label{fig:slater_koster}
\end{figure*}

We fit only the NN-dominated bands near the Fermi level, as shown in Fig.~\ref{fig:slater_koster}, and emphasize that the purpose of this fit is not to reproduce the electronic structure of Ca$_2$RuO$_4$ quantitatively.
Rather, the goal is to obtain physically reasonable hopping amplitudes whose dependence on octahedral geometry generates controlled variations of the resulting hybridization functions across layered-perovskite-like configurations.
In particular, the oxygen-mediated $d$--$p$--$d$ pathway, encoded in the effective hoppings, captures the leading dependence on RuO$_6$ rotations and distortions.
Once the Slater--Koster parameters are fixed, dataset diversity is generated by systematically varying oxygen positions within the RuO$_6$ octahedra, while keeping all electronic parameters unchanged and updating only the local rotation matrices $R^{(i)}$ in Eq.~\eqref{eq:dpd_hop}.

Fig.~\ref{fig:sampling}(c) illustrates the structural sampling
procedure. Starting from a reference symmetric configuration (red atoms),
oxygen atoms are systematically displaced to generate distorted structures. The orientation of each RuO$_6$ octahedron is parameterized
by Euler angles $(\alpha, \beta, \gamma)$ in the $zyz$ convention, which
specify the direction of a reference Ru--O bond. The reference oxygen is
displaced on a spherical patch defined by angular ranges:
\begin{equation}
\theta \in [1.37,\, \pi/2], \quad 
\phi \in [\pi/2 - 0.31,\, \pi/2 + 0.31],
\label{eq:angle_range}
\end{equation}
corresponding to a polar range of approximately $12^\circ$ from the
equatorial plane and an azimuthal wedge of $\pm 18^\circ$ about the
local $y$-axis. This spherical patch is discretized into an
$N_g \times N_g$ uniform grid with $N_g = 32$, yielding
$N_{\mathrm{samples}} = 1024$ distinct configurations.

For each sampled oxygen position, the local rotation matrix $\hat{R}^{(1)}$
for the reference Ru site is constructed from the Euler angles
$(\alpha, \beta, \gamma)$, as illustrated in Fig.~\ref{fig:sampling}(a).
Three Ru sites---$\hat{R}^{(2)}$, $\hat{R}^{(3)}$, and $\hat{R}^{(4)}$---are derived
from $\hat{R}^{(1)}$ by applying the glide plane symmetry operations of the
layered perovskite structure, ensuring that all sampled configurations
preserve the crystallographic symmetry of the parent lattice.
Fig.~\ref{fig:sampling}(c) illustrates this procedure: starting
from a symmetric reference configuration (top), oxygen atoms are
systematically displaced within their allowed angular ranges (middle),
yielding distorted octahedral configurations (bottom, blue atoms) while preserving
the glide plane symmetry.
For each configuration, the oxygen-mediated hopping Eq.~\eqref{eq:dpd_hop}
and direct $d$--$d$ hopping are recomputed from the updated rotation
matrices $\{\hat{R}^{(i)}\}_{i=1}^{4}$. The resulting ensemble of tight-binding
Hamiltonians spans a representative range of octahedral tilts and rotations
relevant to layered ruthenates.

\subsubsection{Training dataset generation}

For each structural configuration, the ground-truth bath parameters serving as regression 
labels are obtained through the following procedure. First, the local Green's function is 
computed via Brillouin zone integration Eq.~\eqref{eq:local_green} in the non-interacting 
limit $\hat{\Sigma}(i\omega_n) = 0$, and the corresponding hybridization function is extracted 
from the Dyson equation:
\begin{equation}
  \hat{\Delta}_{\text{latt}}(i\omega_n) = (i\omega_n + E_{\text{F}}) \hat{I} 
  - \hat{E}_{\text{imp}} - \hat{G}_{\text{loc}}^{-1}(i\omega_n).
  \label{eq:hyb_dyson}
\end{equation}
The calculations are performed at fictitious inverse temperature $\beta = 128$, retaining 
$N_\omega = 512$ fermionic Matsubara frequencies.

The bath parameters $\pmb{\theta}$ are then determined by minimizing the cost function $\chi^2$ defined in Eq.~\eqref{eq:cost_function} using the conjugate gradient algorithm with the convergence criteria specified in Section~\ref{subsec:bath_fitting}. 
Since the machine learning model requires accurately converged solutions as training targets, 
we adopt strict tolerances to ensure high-quality labels.
To mitigate entrapment in suboptimal local minima, multiple random initializations are 
employed for each sample, and the solution yielding the lowest final cost function is retained.

Once the bath parameters have converged, the discrete bath representation defines a fitted 
hybridization function $\hat{\Delta}_{\mathrm{bath}}(i\omega_n; \pmb{\theta})$ via 
Eq.~\eqref{eq:discrete_hyb}. For training the regression model, we use the lattice hybridization 
function $\hat{\Delta}_{\text{latt}}(i\omega_n)$ as input and the optimized bath parameters 
$\pmb{\theta}^* = \{\varepsilon_l, V_{\mu l}\}_{l=1}^{N_{\mathrm{b}}}$ as output labels. 
This pairing ensures that the model learns to predict bath parameters that, when substituted into 
Eq.~\eqref{eq:discrete_hyb}, yield a discrete approximation closely matching the continuous 
target.

The resulting dataset consists of $N_{\text{samples}} = 1024$ input-output pairs 
$\{(\hat{\Delta}^{(s)}_{\text{latt}}, \pmb{\theta}^{(s)*})\}_{s=1}^{N_{\text{samples}}}$, 
where the inputs are symmetry-reduced hybridization functions and the outputs are the 
corresponding symmetry-reduced bath parameters, represented as described in 
Section~\ref{sec:inout_representation}.

\subsection{Non-ML baseline: heuristic initialization}
\label{subsec:heuristic}

To provide a baseline for comparison, we propose a heuristic initialization 
method representing the level of sophistication accessible to a non-expert 
practitioner. 
This method exploits only textbook-level knowledge: the 
spectral representation of the hybridization function and the trace sum 
rule. 
It does not rely on prior DMFT experience, intuition about bath 
energy distributions for specific material classes, or access to previously 
converged calculations. 
This baseline thus represents a fair comparison 
target for evaluating whether machine learning can effectively encode 
expert knowledge into an automated initialization procedure.

\subsubsection{Spectral function and sum rule}

The spectral representation of the hybridization function on the real frequency axis is characterized by
\begin{equation}
A(\omega) = -\frac{1}{\pi} \mathrm{Im}\, \mathrm{Tr}\big[\hat{\Delta}(\omega + i\eta)\big],
\label{eq:spectral_func}
\end{equation}
where $\eta \to 0^+$ is a positive infinitesimal broadening. The trace sum rule provides the total spectral weight:
\begin{equation}
I_{\mathrm{trace}} = \int_{-\infty}^{\infty} A(\omega)\, d\omega.
\label{eq:sum_rule}
\end{equation}
This quantity must be preserved by any physically consistent bath parametrization.

\subsubsection{CDF-based energy sampling}

Rather than relying on peak detection algorithms, which may be sensitive to noise and minor spectral features, we interpret the normalized spectral function as a probability density function (PDF):
\begin{equation}
p(\omega) = \frac{A(\omega)}{\int A(\omega')\, d\omega'}.
\label{eq:pdf}
\end{equation}
The corresponding cumulative distribution function (CDF) is
\begin{equation}
F(\omega) = \int_{-\infty}^{\omega} p(\omega')\, d\omega'.
\label{eq:cdf}
\end{equation}

For $N_{\mathrm{b}}$ independent bath onsite energies (accounting for Kramers degeneracy under time-reversal symmetry), we sample at uniformly spaced quantiles:
\begin{equation}
q_k = \frac{k - 0.5}{N_{\mathrm{b}}/2}, \quad l = 1, \ldots, N_{\mathrm{b}},
\label{eq:quantiles}
\end{equation}
and determine bath onsite energies via the inverse CDF:
\begin{equation}
\varepsilon_k^{\mathrm{indep}} = F^{-1}(q_k).
\label{eq:inverse_cdf}
\end{equation}

\begin{figure*}[h]
 \centering
 \includegraphics[width=0.9\textwidth]{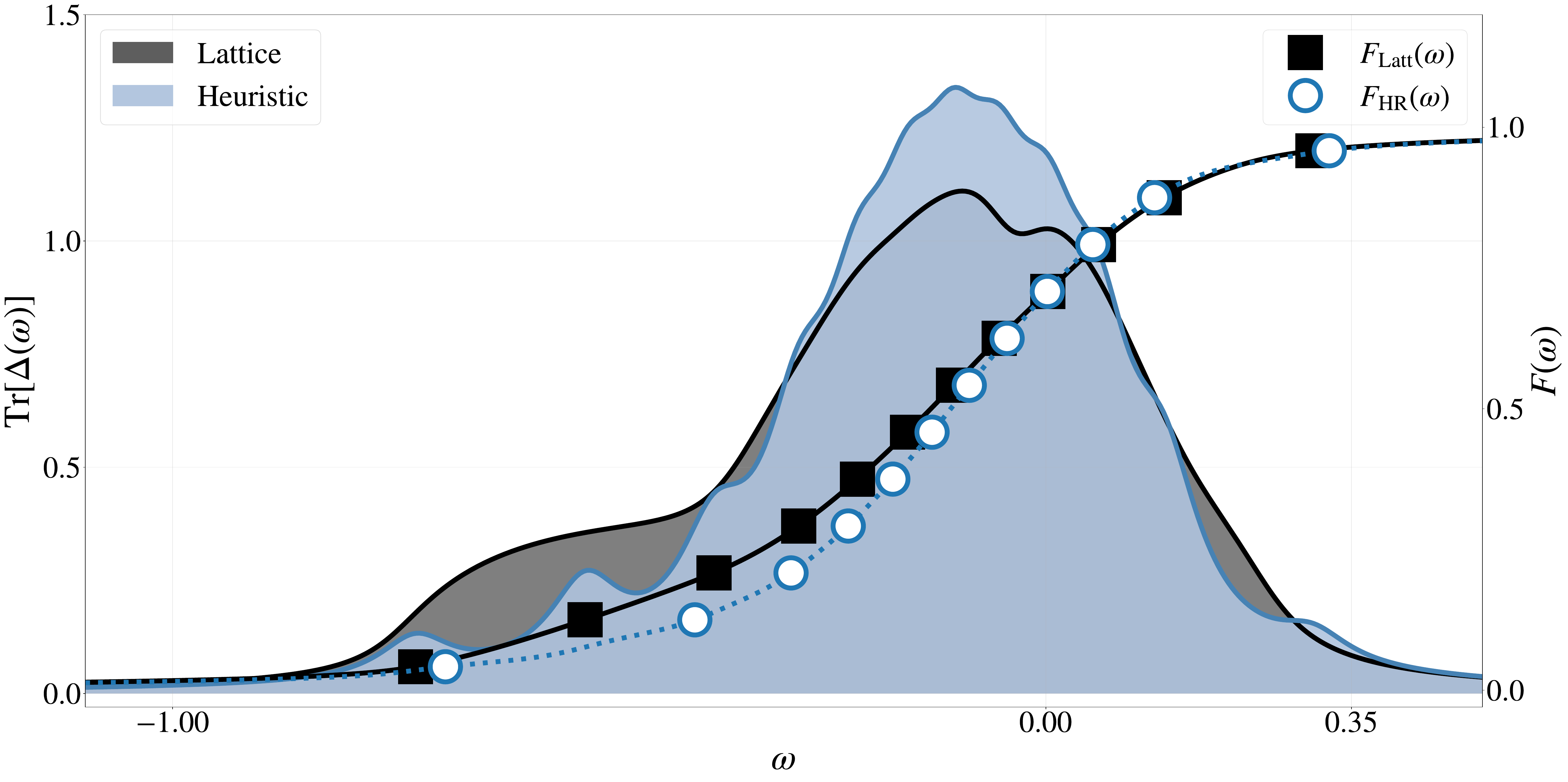}
 \caption{
    Illustration of the heuristic initialization procedure based on cumulative distribution function (CDF) sampling.
    The spectral function $\mathrm{Tr}\big[\hat{\Delta}_{\mathrm{latt}}(\omega)\big]$ from the lattice calculation (grey shaded region) is used to construct the cumulative distribution function $F_{\mathrm{latt}}(\omega)$ (black curve with squares).
    Bath onsite energies (black squares) are determined by inverting the CDF at uniformly spaced quantiles, automatically concentrating bath sites in regions of high spectral weight.
    The heuristic initialization result (blue dotted curve) and its corresponding CDF $F_{\mathrm{HR}}(\omega)$ (blue curve with circles) demonstrate that the method approximately reproduces the overall spectral weight distribution while satisfying the trace sum rule by construction. (For interpretation of the references to color in this figure legend, the reader is referred to the web version of this article.)
    }

\label{fig:cdf_sampling}
\end{figure*}

This procedure automatically concentrates bath sites in regions of high spectral weight. Sharp peaks in $A(\omega)$ produce steep gradients in the CDF, resulting in dense clustering of bath onsite energies near those features. Fig.~\ref{fig:cdf_sampling} illustrates this sampling procedure: the spectral function $\mathrm{Tr}\big[\hat{\Delta}(\omega)\big]$ determines the CDF $F(\omega)$, and the bath onsite energies are obtained by inverting the CDF at uniformly spaced quantiles. The resulting bath onsite energy distribution adaptively reflects the underlying spectral structure, with denser sampling in regions of higher spectral weight.

\subsubsection{Hybridization strength assignment}

Given the bath onsite energies, the hybridization strengths are initialized to approximately reproduce the local spectral weight while satisfying the trace sum rule. For each bath site $k$:
\begin{enumerate}
\item Evaluate the spectral density at the bath onsite energy: $A_k = A(\varepsilon_k)$.
\item Compute the relative site weight: $w_k = A_k / \sum_j A_j$.
\item Assign the spectral weight contribution: $S_k = I_{\mathrm{trace}} \cdot w_k$.
\item Distribute equally among orbitals: $|{V}_{\mu k}|^2 = S_k / N_{\mathrm{orb}}$.
\item Assign random phases: ${V}_{\mu k} = |{V}_{\mu k}| e^{i\theta_{\mu k}}$, where $\theta_{\mu k} \in [0, 2\pi)$.
\end{enumerate}
Fig.~\ref{fig:cdf_sampling} demonstrates that this procedure yields an initial hybridization function (blue shaded) that captures the essential features of the target lattice hybridization function (grey shaded), including the positions and relative heights of the principal spectral peaks.

For systems with time-reversal symmetry, the bath sites form Kramers pairs with degenerate energies $\varepsilon_{2l-1} = \varepsilon_{2l}$. 
Only the even-indexed sites are initialized independently; the odd-indexed sites are determined by the time-reversal symmetry in Section~\ref{sec:trs_constraints}.
This heuristic initialization satisfies several important properties by construction: (i) the trace sum rule is preserved, (ii) time-reversal symmetry constraints are exactly satisfied, and (iii) bath sites are adaptively distributed according to the spectral structure of the lattice hybridization function.

\section{Results}
\label{sec:results}

This section presents numerical results demonstrating the effectiveness of the machine-learning-based initialization strategy for bath fitting in the non-interacting limit. We systematically compare the ML-based approach against the heuristic initialization method described in Section~\ref{subsec:heuristic} across multiple performance metrics: prediction quality of bath parameters, convergence dynamics, robustness against local minima, and scalability with increasing bath size.

\subsection{Prediction quality of bath parameters}
\label{subsec:prediction_quality}

\begin{figure*}[!htbp]
  \centering
  \includegraphics[width=\textwidth]{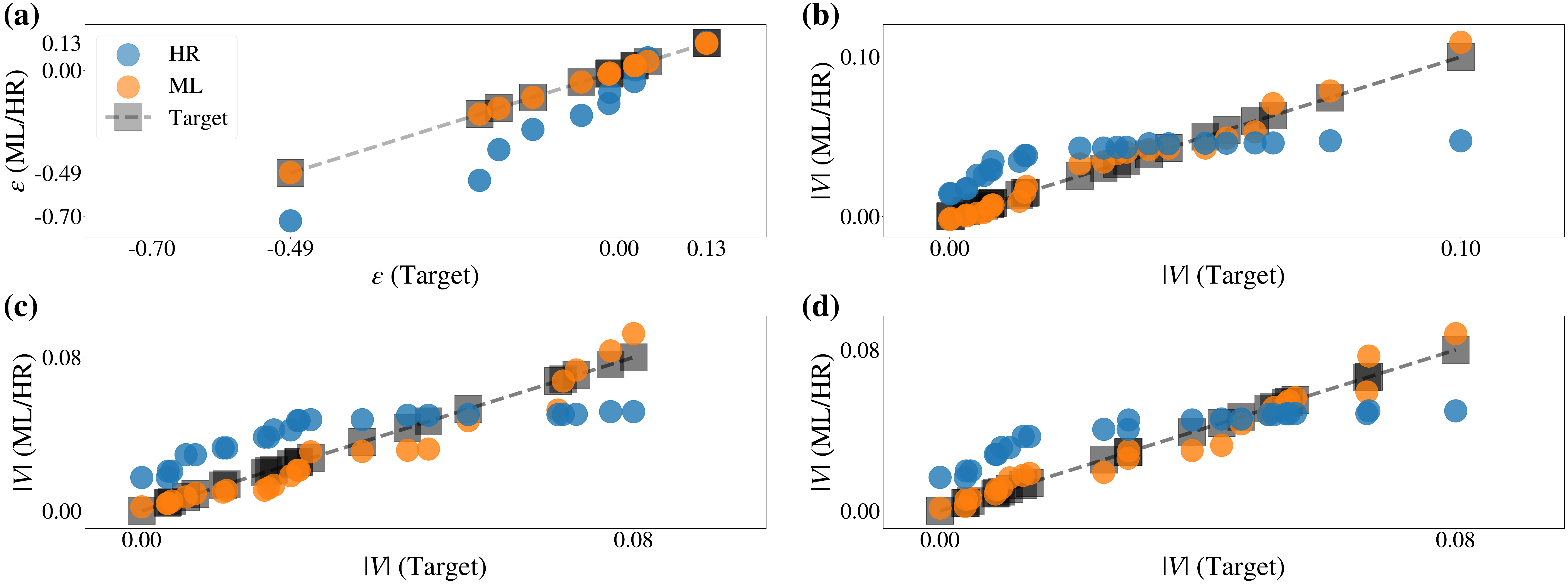}
  \caption{Parity plots comparing predicted bath parameters from machine learning (ML, orange circles) and heuristic (HR, blue circles) initialization against target values (gray squares) for $N_{\mathrm{b}} = 12$. (a) Bath onsite energies $\varepsilon$. (b)--(d) Hybridization strength magnitudes $|{V}_{\mu l}|$ for the three symmetry-inequivalent $j$-effective orbital $\mu$ sectors for all bath sites: (b) $\big|j=\tfrac{3}{2}, m_j=\tfrac{3}{2}\Big\rangle$, (c) $\big|j=\tfrac{3}{2}, m_j=\tfrac{1}{2}\Big\rangle$, and (d) $\big|j=\tfrac{1}{2}, m_j=\tfrac{1}{2}\Big\rangle$. The dashed diagonal lines indicate perfect agreement. The ML predictions consistently lie closer to the diagonal across all parameter types, demonstrating superior prediction accuracy. (For interpretation of the references to color in this figure legend, the reader is referred to the web version of this article.)}
  \label{fig:parity}
\end{figure*}

The primary objective of the ML model is to predict bath parameters that lie closer to the optimal solution than those provided by heuristic initialization. Fig.~\ref{fig:parity} presents parity plots comparing the predicted bath parameters from both ML and HR initialization against the target values obtained from fully converged optimization at $N_{\mathrm{b}} = 12$.

Fig.~\ref{fig:parity}(a) compares the predicted bath onsite energies $\varepsilon_l$. 
The ML predictions (orange circles) cluster tightly along the diagonal, indicating accurate reproduction of the target bath onsite energies across the entire energy range from $\varepsilon \approx -0.49$ to $\varepsilon \approx 0.13$. 
In contrast, the heuristic predictions (blue circles) exhibit substantial scatter, particularly in the intermediate energy regime where multiple bath sites compete for spectral weight.
The hybridization strength magnitudes $|{V}_{\mu l}|$ exhibit orbital-dependent prediction quality, reflecting the varying complexity of the hybridization structure across different $j$-effective states. 
Fig.~\ref{fig:parity}(b-d) display parity plots for the three symmetry-inequivalent orbital sectors: $|j=\tfrac{3}{2}, m_j=\tfrac{3}{2}\rangle$, $|j=\tfrac{3}{2}, m_j=\tfrac{1}{2}\rangle$, and $|j=\tfrac{1}{2}, m_j=\tfrac{1}{2}\rangle$, respectively. 
Across all orbital sectors, the ML predictions demonstrate markedly improved accuracy compared to the heuristic initialization. 

\subsection{Convergence dynamics}

\begin{figure*}[!htbp]
  \centering
  \includegraphics[width=\textwidth]{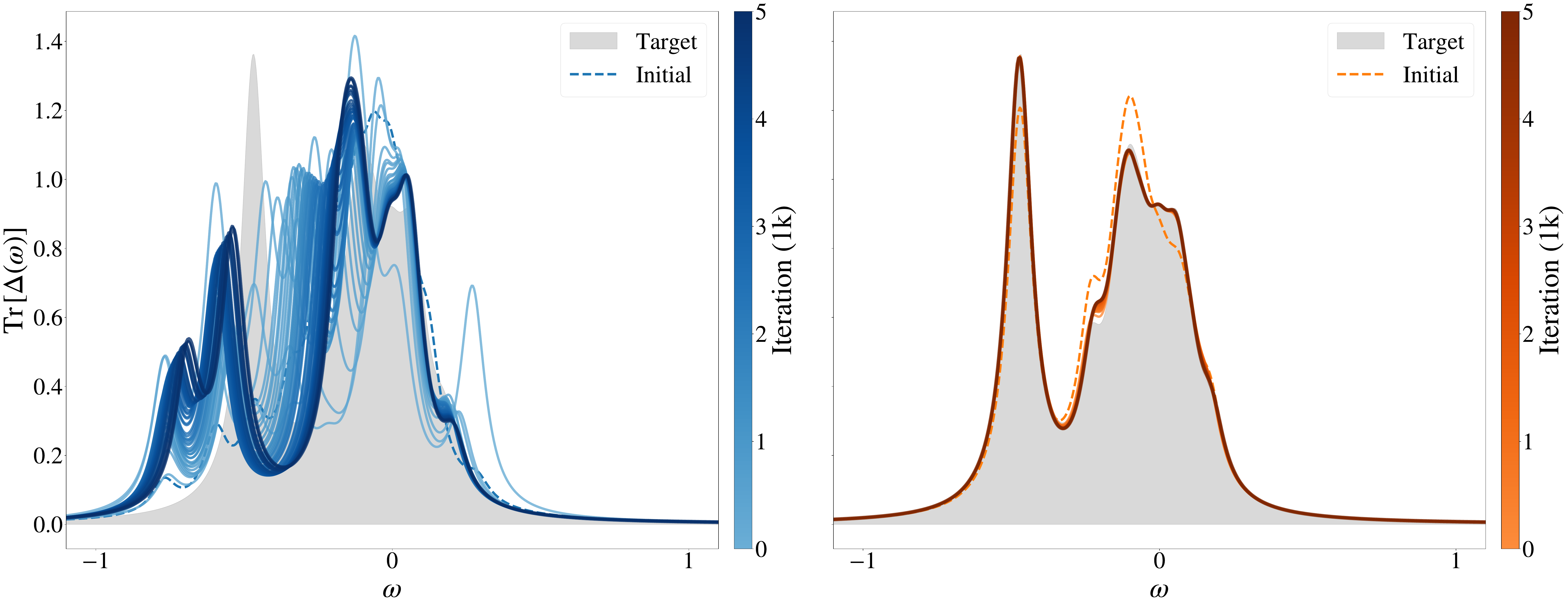}
  \caption{
   Evolution of the fitted hybridization function spectral weight $\mathrm{Tr}\big[\hat{\Delta}(\omega)\big]$ during conjugate gradient optimization for heuristic (HR, left panel, blue) and machine learning (ML, right panel, orange) initialization for $N_{\mathrm{b}} = 12$.
   The gray shaded region represents the target hybridization function.
   Color intensity indicates iteration number, with lighter shades corresponding to earlier iterations (colorbar in units of $10^3$ iterations).
   The ML initialization begins substantially closer to the target and converges within approximately 1000 iterations, whereas the heuristic initialization requires over 5000 iterations to achieve comparable accuracy. (For interpretation of the references to color in this figure legend, the reader is referred to the web version of this article.)
   }
  \label{fig:convergence}
\end{figure*}

The improved prediction quality translates directly into accelerated convergence of the conjugate gradient optimization.
Fig.~\ref{fig:convergence} illustrates the evolution of the fitted hybridization function spectral weight $\mathrm{Tr}\big[\hat{\Delta}(\omega)\big]$ over the course of optimization for both initialization methods.
The left panel shows the trajectory from heuristic initialization, where the initial hybridization function (lightest blue) bears little resemblance to the target (gray shaded region). 
The optimization requires 5000 iterations to bring the fitted spectrum into agreement with the target, traversing a circuitous path through parameter space.

In contrast, the right panel demonstrates that ML initialization places the starting point near the target spectrum. The initial ML-predicted hybridization function already captures the essential features of the target, including the positions and relative weights of the principal spectral peaks. Consequently, the optimization converges within approximately 1000 iterations---a factor of five reduction compared to the heuristic method. This acceleration is particularly significant given that bath fitting is performed at every iteration of the DMFT self-consistency loop; cumulative savings over a typical DMFT calculation can amount to substantial reductions in total computational time.

\subsection{Robustness against local minima}

A central advantage of ML initialization is that it tends to place the starting point inside favorable basins of attraction of the non-convex cost landscape, thereby reducing the risk of convergence to suboptimal local minima. To quantify this robustness, we perform multiple independent conjugate-gradient optimizations for a fixed target hybridization function, applying different random perturbations to both the ML- and heuristic-initialized bath parameters.

\begin{figure*}[!htbp]
  \centering
  \includegraphics[width=\textwidth]{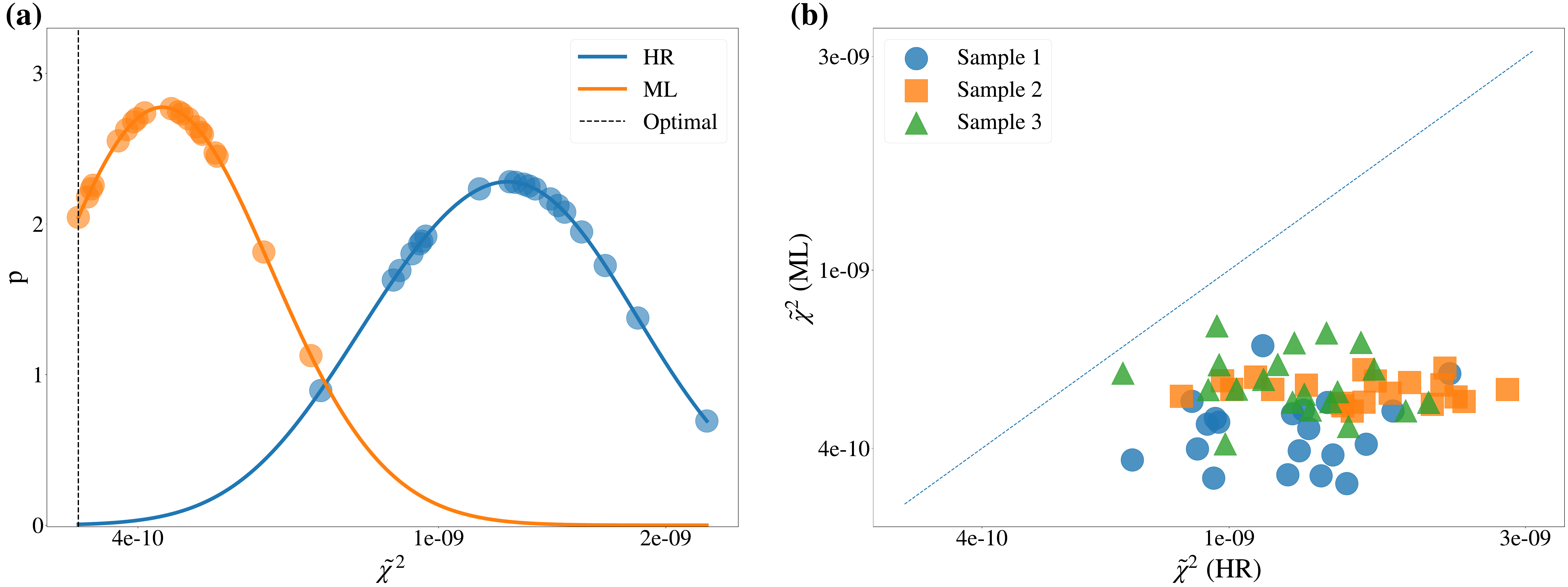}
  \caption{Robustness analysis of ML versus heuristic initialization at $N_{\mathrm{b}} = 18$ (same Euler-angle sample). (a) Probability density $p$ of final converged cost function values $\tilde{\chi}^2$, estimated by Gaussian kernel density estimation in $\log_{10}\tilde{\chi}^2$ space, for heuristic (HR, blue) and machine learning (ML, orange) initialization, constructed from 20 independent optimization runs. Vertical dashed lines indicate the mean values: $\langle \tilde{\chi}^2 \rangle_{\mathrm{ML}} = 4.31 \times 10^{-10}$ (orange) and $\langle \tilde{\chi}^2 \rangle_{\mathrm{HR}} = 1.23 \times 10^{-9}$ (blue).
  The dashed line indicates the ideal limit (exactly zero, attainable as $N_\mathrm{b} \to \infty$).
  (b) Pairwise comparison of converged $\tilde{\chi}^2$ values between HR and ML initialization for three independent samples (three different Euler-angle sets) (different marker shapes).
  Points below the dashed diagonal indicate superior ML performance. (For interpretation of the references to color in this figure legend, the reader is referred to the web version of this article.)}
  \label{fig:robustness}
\end{figure*}

Fig.~\ref{fig:robustness}(a) shows the distribution of converged cost-function $\tilde{\chi}^2(\pmb{\theta}^*_\mathrm{X})$ at $N_{\mathrm{b}} = 18$. The ML-initialized results (orange curve) are tightly concentrated around $\tilde{\chi}^2 \approx 4.31 \times 10^{-10}$, indicating highly consistent convergence to essentially the same minimum. In contrast, the heuristic initialization (blue curve) converges to a broader set of minima, with a mean peak near $\tilde{\chi}^2 \approx 1.23 \times 10^{-9}$ (about three times larger) and a pronounced tail toward higher values.

This systematic advantage can be summarized by the ratio of mean converged costs,
\begin{equation}
  r_{\chi^2} = \frac{\langle \tilde{\chi}^2 \rangle^{}_{\mathrm{HR}}}{\langle \tilde{\chi}^2 \rangle^{}_{\mathrm{ML}}} \approx 2.85,
\end{equation}
which indicates that ML initialization does not merely accelerate convergence to the same solution, but tends to steer the optimizer toward better minima.

To enable a uniform robustness statistic across bath sizes in Section~\ref{subsec:scalability}, we define an optimization run as ``valid'' when its final cost satisfies
\begin{equation}
  \tilde{\chi}^{2}(\pmb{\theta}_{\mathrm X}^*) \le r_\mathrm{tol}\,\tilde{\chi}^{2}(\pmb{\theta}_{\mathrm{target}}^*),
\end{equation}
where $\tilde{\chi}^2(\pmb{\theta}_{\mathrm{target}}^*)$ is the lowest cost observed among our best-performing configurations for the corresponding $\hat{\Delta}_{\mathrm{latt}}$, serving as an empirical proxy for a near-optimal minimum at fixed $N_{\mathrm{b}}$. 
From the analysis at $N_{\mathrm{b}} = 18$, we extract a representative improvement 
factor $r_{\chi^2} \approx 2.85$, which quantifies the typical reduction in initial 
cost achieved by ML initialization relative to the heuristic method. To assess the 
sensitivity of the valid convergence criterion to the choice of threshold, we 
evaluate the valid convergence ratio for four prefactors 
$r_{\mathrm{tol}} \in \{2, 2.5, 3, 3.5\}$, chosen to bracket this empirical value, 
as shown in Fig.~\ref{fig:robustness}(a). Across all tested thresholds, ML 
initialization consistently outperforms the heuristic method, demonstrating that 
the observed robustness advantage is not an artifact of a particular threshold 
choice.

To confirm that the improvement persists across structural configurations, Fig.~\ref{fig:robustness}(b) compares $\tilde{\chi}^2_{\mathrm{HR}}$ (horizontal axis) and $\tilde{\chi}^2_{\mathrm{ML}}$ (vertical axis) for three independent samples at $N_{\mathrm{b}} = 18$. Nearly all points fall below the diagonal, demonstrating that ML initialization consistently yields lower converged costs and reduces sample-to-sample variability.

Finally, we emphasize that the heuristic distribution in Fig.~\ref{fig:robustness}(a) is
itself built from many random-seed initial conditions---effectively a stochastic multistart
of a perturbed heuristic---so its persistently higher converged cost relative to ML cannot be
attributed to a particular or insufficiently randomized choice of heuristic, but reflects the
accuracy of the ML-predicted bath.

\subsection{Scalability with bath size}
\label{subsec:scalability}

As the number of bath sites $N_{\mathrm{b}}$ increases, the dimensionality of the optimization problem grows according to Eq.~\eqref{eq:num_params}, and the cost function landscape becomes increasingly rugged with an increasing number of local minima. This poses a severe challenge for heuristic initialization methods, which lack the ability to adapt to the increased complexity.

\begin{figure*}[!htbp]
\centering
\includegraphics[width=\textwidth]{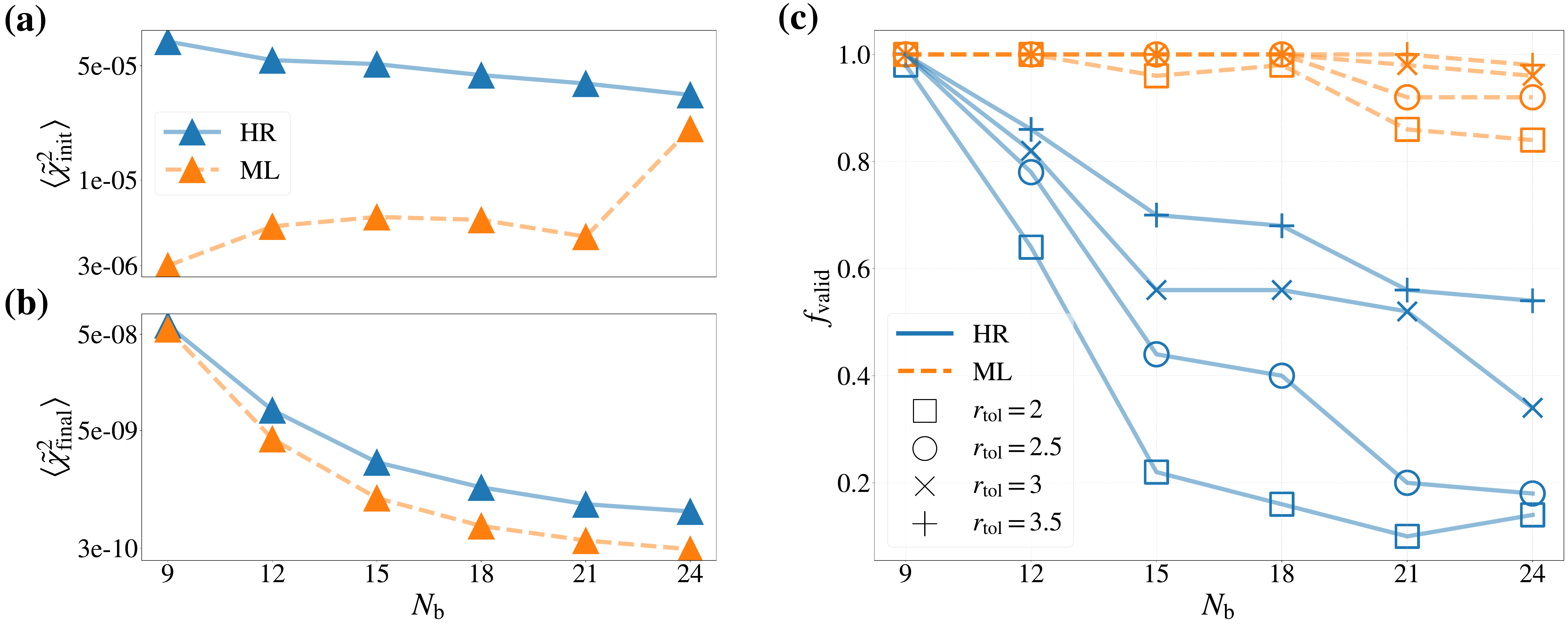}
\caption{
  Scalability analysis of machine learning versus heuristic initialization. 
  Comparison of machine learning (ML) and heuristic (HR) initialization methods as a function of bath size $N_\mathrm{b}$, where $N_\mathrm{b}$ denotes the number of bath sites. 
  (a) shows the mean initial cost function value $\langle \tilde{\chi}^2_{\mathrm{init}} \rangle$ before optimization.
  (b) displays the mean converged cost function value $\langle \tilde{\chi}^2_{\mathrm{final}} \rangle$ after conjugate-gradient optimization, where $\tilde{\chi}^2$ measures the squared deviation between the target and discrete hybridization functions.
  (c) presents the valid convergence ratio $f_{\mathrm{valid}}$, defined as the fraction of optimization runs converging to solutions within a factor $r_\mathrm{tol}$.
}
\label{fig:scalability}
\end{figure*}

Fig.~\ref{fig:scalability}(a) presents the mean initial cost function values $\langle \tilde{\chi}^2_{\mathrm{init}} \rangle$ as a function of bath size for both initialization methods. Across all bath sizes considered, the ML initialization (orange squares) consistently yields lower initial cost function values than the heuristic initialization (blue triangles), demonstrating that the data-driven approach provides bath parameters that lie systematically closer to optimal solutions.
The ML-predicted initial cost remains relatively stable in the range $N_{\mathrm{b}} = 9$ to $N_{\mathrm{b}} = 18$.
At $N_{\mathrm{b}} = 24$, a noticeable increase in $\langle \tilde{\chi}^2_{\mathrm{init}} \rangle_{\mathrm{ML}}$ is observed.
The size of the training dataset is kept fixed for all bath sizes, while the dimensionality of the bath parameter space grows rapidly with $N_{\mathrm{b}}$.
The degradation at larger $N_{\mathrm{b}}$ therefore reflects the increased complexity of the learning problem relative to the available training data, rather than a fundamental limitation of the data-driven approach.
Tests with a larger training set show clear improvement at larger bath sizes.
The main purpose of this work is to demonstrate that even a moderately trained ML model can outperform heuristic initialization in a systematic manner, and that its performance can be further improved by increasing the amount of training data.

While the initial cost provides a measure of proximity to an optimal solution, it does not by itself determine the final outcome of the nonlinear optimization.
The more relevant question is whether the optimization algorithm can reliably converge to high-quality solutions as the bath size increases.
To quantify convergence robustness as a function of bath size, we define the valid convergence ratio
\begin{equation}
  f_{\mathrm{valid}}(N_{\mathrm{b}}) = \frac{N_{\mathrm{valid}}(N_{\mathrm{b}})}{N_{\mathrm{total}}},
\end{equation}
where $N_{\mathrm{valid}}$ counts optimization runs that converge to a cost function value within a factor of $r_{\mathrm{tol}}$ of the best known solution, with $r_{\mathrm{tol}} \in \{2,\, 2.5,\, 3,\, 3.5\}$, and $N_{\mathrm{total}}=50$ is the total number of samples tested at each $N_{\mathrm{b}}$.
This metric captures both convergence success and solution quality.

Fig.~\ref{fig:scalability}(b) displays the mean final (converged) cost function values $\langle \tilde{\chi}^2_{\mathrm{final}} \rangle$ after conjugate gradient optimization as a function of bath size. 
For small bath sizes ($N_{\mathrm{b}} = 9$), both initialization methods achieve comparable final accuracy, with $\langle \tilde{\chi}^2_{\mathrm{final}} \rangle_{\mathrm{ML}} \approx \langle \tilde{\chi}^2_{\mathrm{final}} \rangle_{\mathrm{HR}} \approx 5 \times 10^{-8}$. 
However, as the bath size increases, a systematic separation emerges: the ML initialization leads to consistently lower converged cost function values compared to the heuristic method. 
At $N_{\mathrm{b}} = 24$, the ML-initialized optimization converges to $\langle \tilde{\chi}^2_{\mathrm{final}} \rangle_{\mathrm{ML}} \approx 3 \times 10^{-10}$, whereas the heuristic initialization yields $\langle \tilde{\chi}^2_{\mathrm{final}} \rangle_{\mathrm{HR}} \approx 8 \times 10^{-10}$---nearly a factor of three larger. 
This growing disparity reflects the increasing likelihood that heuristic initialization places the optimization within the basin of attraction of a suboptimal local minimum as the parameter space dimensionality grows. 
The ML initialization, by contrast, systematically identifies favorable basins that lead to higher-quality solutions, demonstrating that the benefits of data-driven initialization extend beyond computational acceleration to encompass improved solution quality in the final converged result.

Fig.~\ref{fig:scalability}(c) presents the valid convergence ratio as a function of bath size for both initialization methods. At the smallest bath size ($N_{\mathrm{b}} = 9$), both methods achieve perfect valid convergence ($f_{\mathrm{valid}} = 1.0$), indicating that the optimization landscape is sufficiently simple to permit reliable convergence from arbitrary starting points. However, as $N_{\mathrm{b}}$ increases, the performance of the two methods diverges dramatically.

The heuristic initialization exhibits a monotonic decline in valid convergence ratio with increasing $N_{\mathrm{b}}$, and for stricter convergence thresholds ($r_{\mathrm{tol}} = 2$), $f_{\mathrm{valid}}$ falls to nearly 0.20 or below for $N_{\mathrm{b}} \geq 15$, indicating that fewer than one in five optimization runs converge to a near-optimal solution.
This behavior reflects the increasing probability of initialization within the convergence of a suboptimal local minimum as the landscape complexity grows. Notably, this degradation occurs even though the initial cost function values for heuristic initialization remain relatively constant with increasing $N_{\mathrm{b}}$, indicating that the quality of the initial guess---not merely its cost function value---determines convergence success in high-dimensional optimization landscapes.

In contrast, the ML initialization maintains $f_{\mathrm{valid}} \geq 0.95$ through $N_{\mathrm{b}} = 18$, and even at $N_{\mathrm{b}} = 21$ and $N_{\mathrm{b}} =24$, it retains high valid convergence ratios of $f_{\mathrm{valid}} \approx 0.80$ and $f_{\mathrm{valid}} \approx 0.95$ for the strictest thresholds $r_{\mathrm{tol}} = 2$ and $r_{\mathrm{tol}} = 2.5$, respectively.
This result demonstrates that while the ML model may not predict optimal initial parameters with the same accuracy at larger bath sizes, the predicted parameters nevertheless lie within basins of attraction that lead to high-quality converged solutions. The robustness of the final convergence outcomes, despite the increased initial cost function values, indicates that the ML initialization successfully identifies favorable regions of parameter space from which gradient-based optimization can reliably reach optimal minima.

The scaling behavior can be understood qualitatively by noting that the number of bath parameters grows as $2N_{\mathrm{b}}(1 + 2N_{\mathrm{orb}}) = 26N_{\mathrm{b}}$ for the six-orbital system considered here, without imposing any symmetry or gauge constraints.  At $N_{\mathrm{b}} = 24$, this amounts to $624$ real parameters in total. Incorporating time-reversal symmetry constraints and fixing the $U(1)$ gauge freedom of each bath Kramers pair reduces this count to $288$ independent real parameters, which constitutes the actual optimization problem solved in practice.
The observation that ML initialization maintains robust convergence even when its direct predictions show some degradation suggests that the model has learned to identify favorable basins of attraction in the cost function landscape rather than merely predicting parameters close to the global minimum.

\subsection{Application to Sr$_2$RuO$_4$}
\label{subsec:sro_application}
\begin{figure*}[!htbp]
  \centering
  \includegraphics[width=\textwidth]{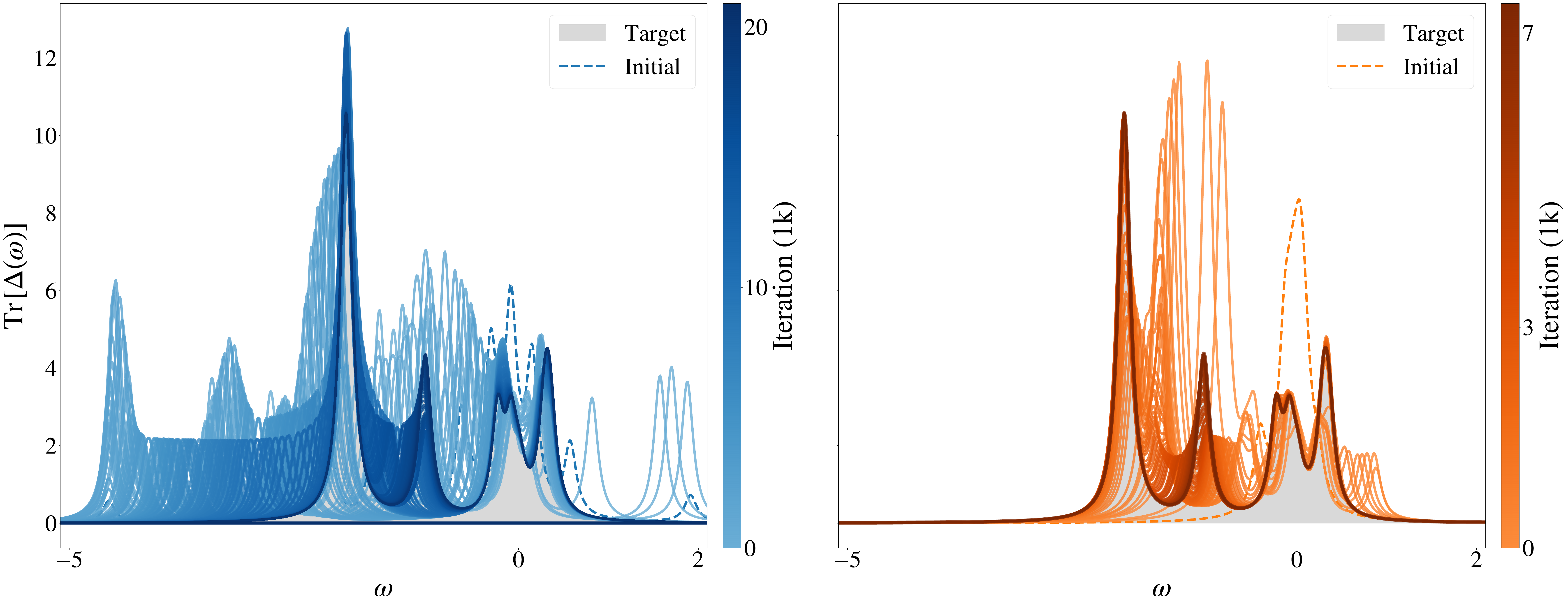}
  \caption{Convergence dynamics of bath fitting for Sr$_2$RuO$_4$ with 
  $U = 2.5$~eV and $J = 0.4$~eV. Evolution of $\mathrm{Tr}\big[\hat{\Delta}(\omega)\big]$ 
  during conjugate gradient optimization for heuristic (left, blue) and 
  ML (right, orange) initialization. Gray shaded region: target hybridization 
  function. Color intensity indicates iteration number (lighter = earlier). 
  Both methods reach identical final solutions; ML initialization requires 
  approximately one-third the iterations. (For interpretation of the references to color in this figure legend, the reader is referred to the web version of this article.)}
  \label{fig:sro_convergence}
\end{figure*}

The preceding analyses established the effectiveness of ML-based initialization in the non-interacting limit.
A natural question is whether these advantages persist when the method is applied to fully interacting DMFT calculations.
To address this, we examine the performance of ML initialization for Sr$_2$RuO$_4$, a canonical strongly correlated metal.

The tight-binding Hamiltonian is constructed from Wannier functions derived from DFT calculations performed with the same computational settings described in Section~\ref{sec:struct_sampling}, providing an accurate representation of the Ru $t^{}_{2g}$ bands near the Fermi level~\citep{Lee2020}.
Local electronic correlations are treated using a Kanamori interaction with $U = 2.5$~eV and $J = 0.4$~eV, parameter values commonly employed in DFT+DMFT studies of ruthenates and supported by comparison with ARPES measurements~\citep{Tamai2019,Lee2020,Kim2023}.
The reference hybridization function is taken from a fully converged DFT+DMFT calculation for Sr$_2$RuO$_4$, obtained using a configuration-interaction-based impurity solver~\citep{Go2015,Go2017}.
To isolate the effect of the initialization strategy, we evaluate bath fitting performance at the final DMFT iteration, where the hybridization function has reached self-consistency.

Fig.~\ref{fig:sro_convergence} presents the evolution of the fitted hybridization function spectral weight $\mathrm{Tr}\big[\hat{\Delta}(\omega)\big]$ during conjugate gradient optimization.
The left panel shows the trajectory from heuristic initialization, while the right panel shows ML initialization.
The gray shaded region indicates the target hybridization function from the DMFT self-consistency condition, and color intensity represents progression through iterations.

The results highlight two main observations. First, the dashed curve in Fig.~\ref{fig:sro_convergence} represents the initial condition. In this particular case, the heuristic initialization is not trapped in a poor local minimum and eventually reaches the correct converged solution, indicating that the heuristic strategy is not intrinsically unreliable. Nevertheless, the subsequent optimization trajectories show a clear advantage of ML initialization: even when the initial spectra do not appear dramatically different by eye, the ML-predicted bath parameters place the fit in a more favorable basin of attraction, leading to a substantially faster approach to the target and convergence within approximately one-third the iterations required by the heuristic method.
Second, and most significantly, this improvement persists despite the model being trained exclusively on non-interacting data. This transferability indicates that the learned mapping captures robust structural relationships between hybridization functions and near-optimal bath parameters that generalize across interaction regimes. The practical implication is substantial: computationally inexpensive non-interacting calculations can be leveraged to accelerate the demanding self-consistency loops of correlated electron simulations.

\section{Conclusion}
\label{sec:conclusion}

This work presents a data-driven initialization strategy for bath fitting in Hamiltonian-diagonalization-based DMFT, motivated by the observation that bath fitting is a highly non-convex optimization problem whose outcome and computational cost depend critically on the starting bath parameters.
We trained a kernel ridge regression model to predict near-optimal discrete bath parameters directly from the target hybridization function, thereby providing physically consistent initial conditions that reduce sensitivity to initialization and improve convergence robustness.

A central ingredient is the physically grounded construction of training data.
Rather than relying on random bath-parameter sampling, we generated hybridization-function/bath-parameter pairs from tight-binding Hamiltonians of layered-perovskite-like ruthenate models across systematically deformed structures, and we used fully converged conventional bath fitting to define the training labels.
To reduce the effective dimensionality and avoid unphysical predictions, we incorporated time-reversal symmetry explicitly in both the feature representation and the output parameterization.
This strategy substantially extends our earlier single-orbital proof-of-concept study~\citep{Kim2024} to realistic multi-orbital models relevant to layered ruthenates.

The numerical benchmarks in the non-interacting limit show that ML initialization consistently lowers the initial fitting error and reduces the number of conjugate-gradient iterations required for convergence over a wide range of structural configurations.
Moreover, the approach remains robust as bath size increases: even when direct prediction accuracy degrades at larger $N_{\mathrm{b}}$, the ML-predicted parameters preferentially fall into favorable basins of attraction, maintaining a high valid-convergence ratio up to $N_{\mathrm{b}}=24$.

Most importantly, we demonstrated that a model trained exclusively on non-interacting data can be applied to an actual interacting DMFT problem.
For Sr$_2$RuO$_4$, solved with an adaptive-truncation impurity solver, the heuristic initialization used here is not trapped in a poor local minimum and eventually reaches the correct converged fit; nevertheless, ML initialization yields a substantially better starting point in practice, leading to a markedly faster convergence trajectory (about one-third the iterations in our test) while preserving the final converged solution.
This transferability across interaction regimes is promising because it shows that computationally inexpensive reference calculations can accelerate demanding self-consistency loops in realistic correlated-electron simulations.

Looking forward, an important direction is to enrich the training distribution beyond
the non-interacting limit, for example by incorporating a representative self-energy
ansatz to generate hybridization functions closer to those encountered throughout an
interacting DMFT cycle, or by assembling DFT-derived hybridization functions from a
broad set of materials to move from the present family-specific initializer toward a
more material-agnostic one.
Building on this, a natural next step is to benchmark the ML initialization against---and to
combine it with---the standard parameter-continuation strategy of reusing a converged bath at
nearby $U$ and $J$ without refitting, across a full interacting DMFT continuation path in which
all inequivalent impurities are treated self-consistently.
It is also promising to extend the present framework beyond single-site DMFT to cluster formulations~\citep{FlorezAblan2025}, where the bath-fitting problem becomes even higher dimensional and robust initialization is correspondingly more critical.
In addition, further improving bath fitting for large numbers of bath orbitals would directly benefit impurity solvers that can treat substantially larger baths (e.g., truncated-diagonalization approaches~\citep{Go2015,Go2017}), since our benchmarks indicate that the relative advantage of ML initialization over heuristic initialization becomes more pronounced as $N_{\mathrm{b}}$ increases.
Such extensions, together with solver- and material-specific adaptations, may further improve initialization quality and enable more reliable automation for high-throughput HD-DMFT studies.

\section*{CRediT authorship contribution statement}
\textbf{Taeung Kim}: Writing -- original draft, Writing -- review \& editing, Software, Methodology, Investigation, Data curation, Visualization, 
Conceptualization, Formal analysis; 
\textbf{Jeongmoo Lee}: Data curation, Software, Writing -- review \& editing;
\textbf{Ara Go}: Writing -- review \& editing, Software, Supervision,
Funding acquisition, Conceptualization.

\section*{Declaration of competing interest}
The authors declare that they have no known competing financial interests or personal relationships that could have appeared to influence the work reported in this paper.

\section*{Acknowledgments}
This work was supported by the National Research Foundation of Korea (NRF) under Grants No. RS-2024-00442775, RS-2023-00256050, RS-2025-00515360, NRF-2023M3K5A1094813, and NRF-2021R1C1C1010429.
The authors thank Choonghyun Kim and Hongkee Yoon for helpful discussions.

\section*{Data availability}
Data will be made available on request.

\bibliographystyle{model1-num-names}

\bibliography{references}

\end{document}